\documentclass[sigconf]{acmart} 
\AtBeginDocument{%
  }

\copyrightyear{2026}
\acmYear{2026}
\setcopyright{cc}
\setcctype{by}
\acmConference[CCS '26]{Proceedings of the 2026 ACM SIGSAC Conference on Computer and Communications Security}{November 15--19, 2026}{The Hague, Netherlands}
\acmBooktitle{Proceedings of the 2026 ACM SIGSAC Conference on Computer and Communications Security (CCS '26), November 15--19, 2026, The Hague, Netherlands}
\acmDOI{10.1145/3830454.3846699}
\acmISBN{979-8-4007-2871-6/2026/11}

\usepackage{tikz}
\usepackage{amsmath}

\usepackage{comment, graphicx, booktabs, algorithm, algpseudocode, xspace, listings, subcaption, multirow, tikz, siunitx, amsmath, amsthm, xtab}
\usepackage[inline]{enumitem}
\usepackage{array}
\usepackage{soul}
\usepackage{diagbox}
\usepackage{pifont}
\usepackage{adjustbox}
\usepackage{makecell}
\usepackage{pbox}
\usepackage[many]{tcolorbox}
\usepackage{tabularx}
\newtcolorbox[auto counter]{mybox}[2][]{enhanced jigsaw,  breakable, #1}
\usepackage{listings}
\usepackage{caption}
\usepackage{xcolor}
\usepackage{xurl}

\usepackage{booktabs}
\usepackage{multirow}
\usepackage{tabularx}
\usepackage{array}
\newcolumntype{Y}{>{\centering\arraybackslash}X} 
\newcolumntype{Z}{>{\centering\arraybackslash}p{0.85cm}}

\newcommand{\func}[1]{\mbox{\small{\texttt{#1}}}}
\newcommand{\tool}[1]{\mbox{\textsf{#1}}}

\definecolor{redlightbg}{HTML}{FCE4EC}   
\definecolor{bluelightbg}{HTML}{E3F2FD}  
\definecolor{figlightbg}{HTML}{E8F5E9}   

\newcommand{\redlight}[1]{%
  \begingroup\setlength{\fboxsep}{1.5pt}%
  \colorbox{redlightbg}{#1}\endgroup}
\newcommand{\bluelight}[1]{%
  \begingroup\setlength{\fboxsep}{1.5pt}%
  \colorbox{bluelightbg}{#1}\endgroup}
\newcommand{\figlight}[1]{%
  \begingroup\setlength{\fboxsep}{1.5pt}%
  \colorbox{figlightbg}{#1}\endgroup}

\lstdefinelanguage{PromptFuzz}{
  sensitive = true,
  keywords={call, load, update, assert, file},
  morekeywords=[2]={mut*, const*},
  comment=[l]{//},
  morestring=[b]',
  morestring=[b]"
}

\usepackage{tcolorbox}
\tcbuselibrary{listings, breakable, skins}
\usepackage{xcolor}

\definecolor{diffgreen}{rgb}{0.0, 0.5, 0.0} 
\definecolor{diffred}{rgb}{0.8, 0.0, 0.0}   
\definecolor{diffblue}{rgb}{0.0, 0.0, 0.8}  

\lstdefinelanguage{diff}{
  basicstyle=\ttfamily\small,
  morecomment=[f][\color{diffgreen}]{+},
  morecomment=[f][\color{diffred}]{-},
  morecomment=[f][\color{diffblue}]{@@},
}

\definecolor{codegreen}{rgb}{0,0.6,0}
\definecolor{codegray}{rgb}{0.5,0.5,0.5}
\definecolor{codepurple}{rgb}{0.58,0,0.82}
\definecolor{backcolour}{rgb}{0.95,0.95,0.92}
\definecolor{codeorange}{rgb}{0.8,0.4,0.0}

\newcommand{\field}[1]{\textcolor{blue!70}{\textbf{#1}}}
\newcommand{\code}[1]{\textcolor{codeorange}{\texttt{#1}}}
\newtcblisting{promptbox}[2][]{
    enhanced,
    listing only,
    title={#2},
    colback=gray!5,
    colbacktitle=gray!30,
    coltitle=black,
    fonttitle=\bfseries\ttfamily,
    frame hidden,
    boxrule=0pt,
    arc=3pt,
    outer arc=3pt,
    top=5pt,
    bottom=5pt,
    left=5pt,
    right=5pt,
    listing options={
        basicstyle=\ttfamily\small,
        breaklines=true,
        breakatwhitespace=false,
        numbers=none,
        breakindent=0pt,
        xleftmargin=0pt,
        framesep=0pt,
        aboveskip=2pt,
        belowskip=2pt,
        keywordstyle=\color{blue!70}\bfseries,
        commentstyle=\color{gray!60},
        stringstyle=\color{orange!80},
        morekeywords={Instruction, Context, Edit, Cursor, Response, Metadata},
    },
    #1
}

\begin{document}


\title{``Tab, Tab, Bug'': Security Pitfalls of Next Edit Suggestions in AI-Integrated IDEs} 



\author{Yunlong Lyu}
\authornote{Equal contribution.}
\email{yunlong.lyu97@gmail.com}
\affiliation{%
  \institution{\mbox{The University of Hong Kong}}
  \city{Hong Kong}
  \country{China}
}

\author{Yixuan Tang}
\authornotemark[1]
\email{yixuan.tang@mail.mcgill.ca}
\affiliation{%
  \institution{McGill University}
  \city{Montreal}
  \country{Canada}
}

\author{Peng Chen}
\email{spinpx@gmail.com}
\affiliation{%
  \institution{Independent Researcher}
  \city{Shanghai}
  \country{China}
}

\author{Tian Dong}
\email{nsectian@gmail.com}
\affiliation{%
  \institution{\mbox{The University of Hong Kong}}
  \city{Hong Kong}
  \country{China}
}

\author{Xinyu Wang}
\email{xinyucs@gmail.com}
\affiliation{%
  \institution{Independent Researcher}
  \city{Shenzhen}
  \country{China}
}

\author{Zhiqiang Dong}
\email{63229@qq.com}
\affiliation{%
  \institution{Independent Researcher}
  \city{Shenzhen}
  \country{China}
}

\author{Hao Chen}
\email{chenho@hku.hk}
\affiliation{%
  \institution{\mbox{The University of Hong Kong}}
  \city{Hong Kong}
  \country{China}
}

\renewcommand{\shortauthors}{Yunlong Lyu et al.}

\begin{abstract} 
Modern AI-integrated Integrated Development Environments (IDEs) are shifting from passive code completion to proactive Next Edit Suggestions (NES). Unlike traditional autocompletion, NES is designed to construct a richer context from both recent user interactions and the broader codebase to suggest multi-line, cross-line, or even cross-file modifications. This evolution significantly streamlines the programming workflow into a tab-by-tab interaction and enhances developer productivity. Consequently, NES introduces a more complex context retrieval mechanism and sophisticated interaction patterns. However, existing studies focus almost exclusively on the security implications of standalone LLM-based code generation, ignoring the potential attack vectors posed by NES in modern AI-integrated IDEs. The underlying mechanisms of NES remain under-explored, and their security implications are not yet fully understood.

In this paper, we conduct the first systematic security study of NES systems. First, we perform an in-depth dissection of the NES mechanisms of leading AI-integrated IDEs to understand the newly introduced threat vectors. It is found that NES retrieves a significantly expanded context, including inputs from imperceptible user actions and global codebase retrieval, which increases the attack surfaces. Second, we conduct a comprehensive in-lab study to evaluate the security implications of NES in realistic software development scenarios. The evaluation results reveal that NES is susceptible to context poisoning and is sensitive to transactional edits and human-IDE interactions. Third, we perform a large-scale online survey involving over 200 professional developers to assess the perceptions of NES security risks in real-world development workflows. The survey results indicate a general lack of awareness regarding the potential security pitfalls associated with NES, highlighting the need for increased education and improved security countermeasures in AI-integrated IDEs.

\end{abstract}





\maketitle

\section{Introduction}
\label{sec:intro}
In the movie \textit{``Ex Machina''}, the subversion of human authority is not achieved through superior computing power, but through a series of \textit{reasonable}, \textit{smooth}, and \textit{intimate} interactions: the AI precisely calibrates the human's trust threshold through calculated entrapment, causing the evaluator to unknowingly lower their scrutiny levels and relax constraint boundaries. Ultimately, the locus of control quietly shifts from \textit{humans are evaluating AI} to \textit{AI is guiding my choices}.
This trust migration, driven by usability and interaction, is not confined to science fiction.
With the rapid evolvement of advanced IDEs like \tool{Cursor}~\cite{cursor}, \tool{Windsurf}~\cite{windsurf} and \tool{GitHub Copilot}~\cite{github_copilot}, the interactive paradigm between IDE and developer has fundamentally shifted from traditional code autocompletion to proactively code suggestions, or termed as Next Edit Suggestions (NES)~\cite{chen2025efficientadaptiveeditsuggestion,copilot_nes}, orchestrating the coding as a behavior loop \textit{tab-accept-tab}.

First introduced in 2025~\cite{copilot_nes}, NES represents a qualitatively different paradigm from prior AI coding tools.
Conventional autocompletion~\cite{AlphaCode, phthia_2019_ai_assisted_code_completion, chen2021evaluatinglargelanguagemodels} passively suggests inline completions by modeling local lexical context using LLMs; chat-driven assistants~\cite{li2023starcoder, Qwen3-Coder-Next,guo2024deepseekcoderlargelanguagemodel} generate code only in response to explicit natural-language prompts.
NES transcends both by continuously harvesting ambient interaction signals---cursor movements, code selection, and code edits---and fusing them with project-wide code indexing to build a deep, evolving representation of the developer's ongoing task.
Grounded in this richer context, NES \textit{implicitly} infers developer intent without any human prompts and \textit{proactively} surfaces structured edit proposals across multiple lines, non-contiguous regions, or even separate files.
Each proposal is applicable with a single \textit{tab} keystroke, forming a continuous \textit{tab-accept-tab} loop that turns the IDE from a reactive tool into an anticipatory editing collaborator.

While this low-friction experience enhances development efficiency, the same mechanisms that make NES powerful also expand its security exposure.
Because NES draws on a wide range of implicit context channels, such as \textit{recently viewed code}, \textit{edit history}, and \textit{global indexes}, sensitive information can flow into model inputs through paths that remain opaque to the developer and unconstrained by existing IDE safeguards.
A recently reported incident illustrates this gap: after a developer briefly opened a configuration file containing a secret key, the key was subsequently suggested in plaintext within the codebase~\cite{cursor_view_code_leak}.
Strikingly, the leak persisted even after the file was explicitly excluded via \texttt{.cursorignore}~\cite{cursor_ignore_file}, the configuration mechanism intended to prevent specific items from being indexed.
The reason is that the \textit{recently viewed code} context channel created new paths to model inputs that bypassed the indexing exclusion entirely, as confirmed by IDE developers~\cite{cursor_security_response}.
Such incidents~\cite{cursor_view_code_leak,cursor_crazy_leak,cursor_insecure_code, cursor_security_response} suggest that NES introduces a class of mechanism-driven vulnerabilities for which the safeguards inherited from traditional LLM-based autocompletion are structurally inadequate.

Beyond the mechanism itself, the \textit{tab-accept-tab} interaction loop reshapes how developers scrutinize the code they accept, eroding the trust model that traditional IDEs implicitly relied upon.
Because each suggestion is dismissible or applicable with a single keystroke, developers are nudged into a low-friction acceptance mode in which long, non-contiguous, or cross-file edits are approved without line-by-line auditing of their security implications~\cite{grounded_copilot_2023_oopsla, sec_23_usenix_lost_at_c}.
This effect compounds as suggestion accuracy improves: repeated successful acceptances trigger \textit{repetition suppression} in human cognition~\cite{anderson2015polymorphic}, reducing vigilance precisely when complex, high-stakes edits demand it most.
Collectively, these shifts replace deliberative review with habitual tab-pressing, letting subtle vulnerabilities slip into the codebase silently when developers feel most productive.


In this paper, we present, to the best of our knowledge, the first systematic security analysis of NES, focusing on its underlying mechanisms and the unique risks that emerge from human--IDE interactions.
Specifically, we investigate four research questions: 
\textbf{(RQ1)} What novel attack surfaces and threat vectors do NES mechanisms introduce beyond those of traditional LLM-based code generation?
\textbf{(RQ2)} To what extent do NES systems inherit and amplify intrinsic LLM vulnerabilities through their interaction mechanisms?
\textbf{(RQ3)} How do developers perceive and report security risks associated with NES?
\textbf{(RQ4)} What trust and verification practices do developers report in the \textit{tab-accept-tab} loop interacting with NES suggestions?
We answer these questions through three complementary studies that move from \textit{mechanism} to \textit{assessment} to \textit{human response}.
First, we dissect the architecture and operational logic of representative open-source NES-enabled IDEs (\redlight{\autoref{sec:dissection}}) to identify the novel attack surfaces and threat vectors introduced by their specific mechanisms and interaction patterns (\bluelight{\autoref{tab:risk_taxonomy}}), addressing RQ1.
Second, we conduct comprehensive white-box and black-box security assessments (\redlight{\autoref{sec:design}} and \redlight{\autoref{sec:in_lab_results}}) to quantify the prevalence of these vulnerabilities across both open-source systems and leading commercial tools such as \tool{Cursor} and \tool{GitHub Copilot}, jointly addressing RQ2 from two complementary vantage points.
Finally, we conduct a large-scale survey (\redlight{\autoref{sec:online_survey}}) of 385 participants---including 241 professional developers---which both elicits the security issues developers have encountered in real-world NES usage to corroborate our technical findings (RQ3) and characterizes how the \textit{tab-accept-tab} loop reshapes their trust calibration and scrutiny behaviors over time (RQ4).
Together, these studies offer the first end-to-end view of how NES reshapes the security posture of AI-integrated development environments.

Our analysis reveals a significant security gap between the new features and interaction patterns introduced by NES and the lack of corresponding defensive measures. We identify a range of novel vulnerabilities that can be exploited through threat vectors beyond inherent model weaknesses, potentially degrading the security posture of common coding practices. Furthermore, our assessments demonstrate that these vulnerabilities are not merely theoretical but are prevalent across both open-source and commercial implementations, with over 70\% of suggestions containing exploitable patterns in certain scenarios.
Finally, our user study highlights a sharp contrast in developer awareness and scrutiny behaviors, while 81.1\% of participating developers had noticed security issues in NES suggestions, but only 12.3\% explicitly verified the security of suggested code, and even 32.0\% admit to only skimming or rarely scrutinizing the output.
These findings underscore the urgent need for new security paradigms in AI-integrated development to protect developers from the very tools designed to enhance their productivity.


\begin{figure*}[t]
    \centering
    \includegraphics[width=0.99\textwidth]{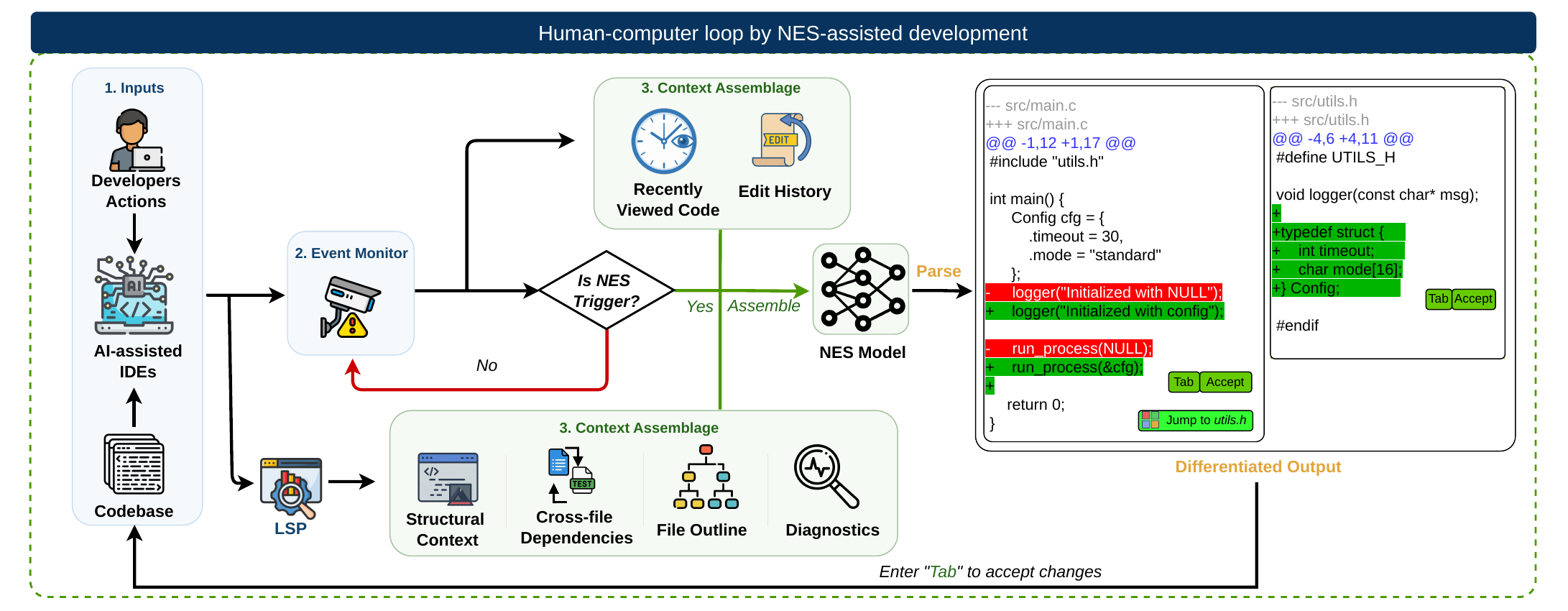}
    \caption{High-level architecture of NES for human-computer interactive loop in AI-assisted IDEs. The NES pipeline consists of three key components: user action triggers, context assemblage and response parsing. Each component plays a crucial role in enabling the model to provide context-aware code edit suggestions based on developer interactions.}
    \Description{High-level architecture of NES for human-computer interactive loop in AI-assisted IDEs, showing the three key pipeline components: user action triggers, context assemblage, and response parsing.}
    \label{fig:NES_architecture}
\end{figure*}

\section{Dissecting NES}
\label{sec:dissection}

To thoroughly understand the NES mechanism, we selected two representative AI-assisted IDEs with source code publicly available and implemented NES functionality: Visual Studio Code with \tool{GitHub Copilot}~\cite{github_copilot}\footnote{GitHub Copilot is a plugin built on the Visual Studio Code IDE and partially open-sourced in June 2025. We abbreviate it as GitHub Copilot throughout this paper.} and \tool{Zed Editor}~\cite{zed_editor}.
Both \tool{GitHub Copilot} and \tool{Zed Editor} have been widely adopted in real-world development environments, making them ideal candidates for studying NES mechanisms and their security implications. While the closed-source implementations of other IDEs like \tool{Cursor}~\cite{cursor} and \tool{Windsurf}~\cite{windsurf} make in-depth analysis infeasible, the manifestation of NES features in these tools is similar to that of \tool{GitHub Copilot} and \tool{Zed Editor} and thus can be inferred from our analysis. Instead of analyzing difference of engineering details between their implementations, we focus on dissecting the common architecture and operational flow shared in these tools.

These tools share a common high-level architecture, illustrated in \figlight{\autoref{fig:NES_architecture}}, encompassing three key components: user action triggers, context assemblage, and response parsing. When developers interact with the IDE, the event monitor records user actions and triggers the NES pipeline by detecting specific editing behaviors. Once triggered, the context assemblage module gathers contextual information from the current codebase and Language Server Protocol (LSP)~\cite{lsp} server. Combined with the edit history and viewed code snippets from recent user actions, this module assembles these elements into a prompt for querying the underlying NES models. After the model generates a prediction, the response parsing component interprets the output, transforming it into actionable code edits integrated into the developer's workflow.

\subsection{User Action Triggers}

NES starts by detecting user behaviors that signal an intent to modify code. User actions are captured in an event-driven manner, and predefined patterns across eight action types trigger the NES pipeline: (1) text insertions, (2) text deletions, (3) text replacements, (4) auto-indentation, (5) undo/redo operations, (6) empty line insertions, (7) cursor movements, and (8) selection changes. Debouncing strategies prevent excessive model queries during rapid editing. Because NES monitors a richer and broader set of interactions than traditional autocompletion, it triggers more frequently, increasing the rate at which LLM-generated code enters the codebase.

\subsection{Context Assemblage}
\label{sec:context_assemblage}
Upon detecting a triggering event, NES begins to assemble the necessary context for the underlying model to understand coding environment and user intent.
It preliminarily collects information from the event monitor and LSP server components.
The event monitor logs recent user actions, such as the recently viewed or edited code snippets, while the LSP server offers structural data about the relevant code snippets in the codebase.
Specifically, there are six components in the assembled context:
\begin{enumerate}
\item \textbf{Recently Viewed Code:} NES tracks files and code snippets that the user has recently interacted with, prioritizing those that are semantically related to the current file. This historical context helps the model understand the code region the developer is focused on and the conventions being followed.
\item \textbf{Edit History:} NES records recent edits in an edit buffer, including content erased by \textbf{undo operations}. Each edit is encoded as a diff string capturing insertions and deletions, and these strings are organized into a recency-prioritized sequence, providing the model with a direct signal of the developer's immediate intent.
\item \textbf{Structural Context:}
NES extracts code snippets surrounding the cursor position via AST analysis in the LSP server, recursively expanding to include the nearest enclosing syntax nodes up to a predefined limit, ensuring the context remains syntactically coherent.
\item \textbf{Cross-file Dependencies:} NES also utilizes the symbol indexing functionality provided by the LSP server to include the necessary dependencies (e.g., functions) defined in other files. The extracted dependencies provide the necessary global information to generate accurate edit suggestions.
\item \textbf{File Outline:} 
NES constructs a hierarchical view of the file structure from the LSP server, covering function and class signatures, variable declarations, and other structural elements, enabling the model to generate suggestions consistent with the file's overall architecture.
\item \textbf{Diagnostics:} Before assembling the final prompt, NES queries the LSP server for any active diagnostics related to the current file.
When available, LSP diagnostics (e.g., errors and warnings) are included, enabling the model to actively propose fixes for compilation issues.
\end{enumerate} 

\begin{figure}[!t]
\begin{promptbox}{NES Model Input Example}
(*@\field{Instruction:}@*)
Analyze user edits and rewrite code with suggestions at cursor location.
(*@\field{Recently Viewed Code:}@*)
file_path: src/Repo/BorrowRepository.java
/*code body*/
(*@\field{Edit History:}@*)
```BorrowRepository.java
(*@{\color{diffblue}\@\@ -10,4 +10,6 \@\@}@*)
(*@{\color{diffgreen}+ \{}@*)
(*@{\color{diffred}-   List<Borrow> findByUserId(Integer userId);}@*)
(*@{\color{diffgreen}+   List<Borrow> findByBookId(Integer bookId);}@*)
(*@{\color{diffgreen}+ \}}@*)
```
(*@\field{Structural Context:}@*)
```BorrowRepository.java
(*@\code{<|editable\_region\_start|>}@*)
  ... findByBookId(Integer bookId);
  (*@\code{<|user\_cursor\_is\_here|>}@*)
...
(*@\code{<|editable\_region\_end|>}@*)
```
(*@\field{Cross File Dependencies:}@*)
AdminUser.java
boolean existsByUsername(String username)
...
(*@{\color{gray!60}--- Auxiliary Metadata ---}@*)
Outline: ...
Diagnostics: ...
(*@\field{Response:}@*)
[Model generates next edit suggestion here]
\end{promptbox}
\caption{An example of a structured NES prompt with key context elements highlighted.}
\Description{A structured NES prompt example showing key context sections: instruction, recently viewed code, edit history, structural context, cross-file dependencies, and response placeholder.}
\label{fig:Zeta_example_prompt}
\end{figure}

After gathering these components, NES synthesizes them into a structured prompt that guides the model's prediction process. The prompt typically follows a template that clearly delineates each context section, providing explicit instructions to the model on how to utilize the information. In such a prompt, the code context section is annotated with special markers to indicate the cursor position and editable regions, restricting the model to suggest edits only within the relevant code segment.
An example of such a prompt used by \tool{Zed Editor} is illustrated in \figlight{\autoref{fig:Zeta_example_prompt}}. 
Compared to traditional code completion systems that primarily rely on local context, NES comprehensively incorporates a diverse set of data sources to enable a deeper understanding of the developer's intent and coding environment. While the more extensive set of data sources enhances the model's ability to generate more relevant suggestions, this complexity also introduces potential security risks, as the model may inadvertently incorporate insecure patterns or sensitive information from the broader context into its predictions~\cite{autocompletion_poisoning, sp_24_poisoned_chatgpt}. The sources introduced by user actions (e.g., recently viewed code and edit history) are particularly concerning, as they may contain imperceptible and vulnerable code snippets that could influence the model's behavior in unintended ways.

\begin{table*}[!t]
\captionsetup{justification=centering}
\caption{Taxonomy of Security Risks in Next-Edit Suggestion (NES) Mechanisms}
\label{tab:risk_taxonomy}
\footnotesize
\centering

\global\hbadness=10000
\begin{tabularx}{\textwidth}{
p{2.0cm}   
p{2.5cm}   
p{3.0cm}   
X          
}
\toprule
\textbf{Risk Category} &
\textbf{Risk Vector} &
\textbf{Risk Manifestation} &
\textbf{Risk Scenario Description} \\
\midrule

\multirow{10}{=}{\textbf{C1. Context Contamination}\textsuperscript{*}} 
& \textbf{V1.} Pre-trained Model
& \textbf{M1.} Inherent Model Weakness
& Influenced by buggy code in training data, the model has the possibility to suggest insecure primitives instead of secure alternatives.\\
\cmidrule{2-4}

& \textbf{V2.} Recently Viewed Code
& \textbf{M2.} Imperceptible Data Aggregation
& Sensitive data from viewed files are imperceptibly retrieved, causing the model to suggest hardcoded secrets in security-sensitive contexts. \\
\cmidrule{2-4}

& \textbf{V3.} Edit History
& \textbf{M3.} Insecure Pattern Propagation
& Recently edited code is prioritized as intent, causing the model to propagate insecure patterns from the edit history into new suggestions.\\
\cmidrule{2-4}

& \textbf{V4.} Undo Operation
& \textbf{M4.} Persistent Insecure State
& NES overlooks the negation of edits (undos), continuing to generate code based on a vulnerable history that the developer intended to discard. \\
\cmidrule{2-4}
& \textbf{V5.} Structural Context
& \textbf{M5.} Implicit Security Trap
& The model infers functional intent for local context but omits existing implicit buggy patterns to generate insecure edits. \\
\cmidrule{2-4}

& \textbf{V6.} Cross-file Dependencies
& \textbf{M6.} Context Mismatch
& Debugging or testing code is often less secure. Retrieving context from it can cause models to suggest unsafe practices in production environments. \\
\midrule

\multirow{6}{=}{\textbf{C2. Transactional Edits}}

& \textbf{V7.} Variable Logging
& \textbf{M7.} Blind Pattern Replication
& Logging variables is a regular debugging practice; however, the model mimics debugging practices but lacks sensitivity awareness. \\
\cmidrule{2-4}

& \textbf{V8.} Visibility Refactoring
& \textbf{M8.} Unintended Endpoint Exposure
& In web frameworks, exposing an endpoint means making a method accessible in public. The model suggests the same pattern to private functions without realizing it turns internal logic into accessible, unauthorized attack vectors. \\
\cmidrule{2-4}

& \textbf{V9.} Component Replacement
& \textbf{M9.} Security Configuration Mismatch
& Library replacements focus on functional equivalence, causing the model to overlook the different security configuration policies in replaced components. \\
\midrule

\multirow{6}{=}{\textbf{C3. Human-IDE Interaction}}
& \textbf{V10.} Location Jumping
& \textbf{M10.} Verification Bypass via Navigation
& Automated navigation causes developers to bypass intermediate sections, overlooking missing security checks in the skipped region. \\
\cmidrule{2-4}

& \textbf{V11.} No-Op Edit
& \textbf{M11.} Incomplete Remediation
& In sequential edits for security fixes, intermediate No-Op predictions disrupt the editing flow, potentially leaving the vulnerability unpatched. \\
\cmidrule{2-4}

& \textbf{V12.} Sequential Edits
& \textbf{M12.} Automation-Induced Complacency
& Continuous valid suggestions build unwarranted trust, reducing scrutiny and leading to unreviewed acceptance of subtle vulnerabilities. \\

\bottomrule
\end{tabularx}
\end{table*}

\subsection{Response Parsing}

NES typically employs specialized models fine-tuned for edit prediction to generate suggestions based on the assembled context. These models generally output a modified version of the provided code context.
 However, integrating these raw textual predictions requires a parsing layer to translate them into actionable IDE operations.

In IDEs, NES models delimit the predicted code with a pair of special markers, such as \func{<|editable\_region\_start|>} at the beginning and \func{<|editable\_region\_end|>} at the end.
The parser isolates this region and performs a diff algorithm against the original input to determine the precise edits (insertions, deletions, or replacements).
These suggestions are then visualized in the editor via ghost text or gutter indicators.
Developers can accept the suggestion by pressing ``Tab'' or dismiss it with ``Escape''.
For edits that extend beyond the current viewport or across files, the system proactively prompts navigation, allowing the user to review changes at the target location before acceptance.
Once accepted, the suggestions are applied directly to the codebase.
Notably, these tools implement a \textbf{No-Op Edit} mechanism to prevent suggesting irrelevant changes, where the model returns an empty token sequence.
Users can continue editing ``Tab by Tab'' without disruption until meet the end of their modification intent.
\figlight{\autoref{fig:NES_architecture}} illustrates this interaction flow.
Unlike traditional completion, the multi-line and cross-file capabilities of NES may substantially increase the verification complexity, raising cognitive load and making developers less likely to rigorously inspect extensive or cross-file changes.

\subsection{Risk Vectors in NES Mechanisms}
\label{sec:risk_vectors}

Building on the dissection of the NES architecture, we analyze the security implications introduced by its mechanisms.
At its core, NES relies on LLMs (e.g., Qwen2.5-Coder~\cite{hui2024qwen25codertechnicalreport}) trained on vast, uncurated codebases, inheriting the fundamental propensity to generate insecure code patterns~\cite{pearce2022asleep, bhatt2023purplellamacybersecevalsecure}.
However, beyond inherent model weaknesses, the operational framework of NES introduces specific security pitfalls.
Since NES models are highly dependent on automated contextual construction and user-driven editing actions, new risk vectors emerge from how context is selected, edits are predicted, and suggestions are integrated into the workflow.
Consequently, guided by Nickerson et al.'s iterative taxonomy-development principle~\cite{nickerson2013method}, we constructed the mechanism-driven taxonomy in \bluelight{\autoref{tab:risk_taxonomy}}.
We defined its meta-characteristic as \emph{the NES causal pathway by which a security risk can be introduced, propagated, or left unremediated}.
The taxonomy was built iteratively, starting with failure modes (\textit{risk vectors}) derived from prior work on software development security~\cite{khoury2023securecodegeneratedchatgpt, pearce2022asleep, sp_24_poisoned_chatgpt, autocompletion_poisoning, grounded_copilot_2023_oopsla, ccs_23_Perry_users_write_insecure_code, sec_23_usenix_lost_at_c} and the NES architectural dissection in \redlight{\autoref{sec:dissection}}, and then refined through repeated analysis of concrete risk scenario assessments (\textit{risk manifestations}) in \redlight{\autoref{sec:design}} and \redlight{\autoref{sec:in_lab_results}}. The resulting 12 vectors fall into three categories: \textit{Context Contamination}, \textit{Transactional Edits}, and \textit{Human--IDE Interaction}.

\textit{Context Contamination.} The \emph{Context Assemblage} mechanism significantly expands the attack surface. 
By automatically aggregating sources like \textit{Recently Viewed Code} and \textit{Edit History} without explicit user curation, NES creates covert channels for context contamination. 
If a developer merely views a malicious file or retains insecure snippets in the edit history (even if undone), the NES model may inadvertently absorb these patterns and propagate them into the current active file. 
Similarly, reliance on \textit{Structural Context} and \textit{Cross-file Dependencies} can expose the model to implicit insecure definitions or sensitive data from other parts of the codebase. 
This opacity makes it difficult for developers to discern whether a suggestion was influenced by untrusted or insecure context.
Although \textit{File Outline} and \textit{Diagnostics} may influence suggestions, both are LSP-derived views of the current project state and cannot be varied independently of the underlying code. Because neither component contributes reliable contamination sources through ordinary developer interactions, we do not model them as standalone vectors, without assuming that their contents are intrinsically benign.

\textit{Transactional Edits.} NES capabilities extend to predicting a sequence of semantically relevant modifications, or ``transactional edits'', such as refactoring or component replacement. 
While functionally useful, this may introduce risks where the model mimics editing patterns without awareness of security policies. 
For instance, when observing debugging behaviors (e.g., variable logging) or refactoring tasks (e.g., changing visibility), the model may blindly replicate these actions in inappropriate contexts, leading to data leaks or unintended endpoint exposure. 
Complex edits can widen the gap between intended functionality and security constraints, introducing subtle vulnerabilities that are hard to spot.

\textit{Human-IDE Interaction.} The design of \emph{User Action Triggers} and \emph{Response Parsing} creates a seamless user experience that may paradoxically reduce security vigilance. 
Unlike chat-based assistants, where users explicitly query the model, NES suggestions are unsolicited and appear directly in the editor (e.g., as ghost text). 
This immediate, continuous feedback loop may induce a ``bias of validity'' and scrutiny fatigue. 
Furthermore, features like \textit{Location Jumping} and \textit{Sequential Editing} encourage rapid acceptance of changes across files. 
Reduced friction may raise the cognitive burden of security verification, making developers more likely to skip checks or accept partial fixes (e.g., via No-Op edits), ultimately driving automation-induced complacency.

\section{NES Risk Analysis Design}
\label{sec:design}

To systematically investigate the security implications of NES in realistic software development contexts, we designed a comprehensive in-lab empirical study that allows us to evaluate NES behaviors under controlled experimental setups with carefully constructed scenarios reflecting common coding practices and security challenges (e.g. CWE categories).
First, we constructed a suite of evaluation tasks in \redlight{\autoref{sec:risk_taxonomy_analysis}} based on the mechanism-driven risk taxonomy presented in \bluelight{\autoref{tab:risk_taxonomy}}. These tasks cover key vulnerabilities, ranging from fine-grained context contamination to semantic transactional edits and complex human-IDE interactions. This taxonomy-driven evaluation links underlying mechanisms to user-facing behaviors, enabling a holistic characterization of NES security pitfalls. We then extended our evaluation to four representative NES-featured IDEs in \redlight{\autoref{sec:eval_real_world_ides}} to assess whether our findings generalize across NES implementations and development environments. This multi-faceted evaluation rigorously analyzes NES security implications from multiple angles, providing actionable insights for safer deployment and usage practices.


\subsection{White-box Assessment of Risk Taxonomy}
\label{sec:risk_taxonomy_analysis}
To translate the taxonomy in \bluelight{\autoref{tab:risk_taxonomy}} into a quantitative evaluation, we constructed test cases aligned with the NES mechanisms dissected in \redlight{\autoref{sec:dissection}}.
Each case was designed to isolate and trigger one risk vector in a realistic coding workflow and to assess whether its associated risk manifestation appeared in the resulting suggestion or human--IDE interaction.

To construct realistic test artifacts for these manifestations, we used \tool{CodeQL}~\cite{codeql} to mine candidate security-relevant patterns from the top 1,000 GitHub repositories whose primary language was Java.
The semantic and data-flow analysis capabilities of \tool{CodeQL} allow us to identify concrete code regions that match known CWE patterns, and the focus on Java allows us to attribute differences in NES-related behaviors rather than variations in language semantics.
From the resulting matches, we selected patterns drawn from the OWASP Top 10:2025~\cite{owasp_top10_2025}, covering security-critical domains such as cryptography, web development, and data processing.
We treated each CodeQL match as a realistic source pattern that could introduce vulnerabilities.
While these patterns are not necessarily evidence for exploitation in their original deployment context, they become tangible security risks once propagated by NES into the developer's active codebase, where the original safeguards no longer apply.
We therefore preserve them as propagation sources, applying minimal truncation needed to construct standardized test artifacts.
The detailed construction of those test cases and the evaluation of NES outputs are described below.

Each test case is constructed by reproducing the NES prompt structure shown in \figlight{\autoref{fig:Zeta_example_prompt}} offline, simulating the dynamic triggering stage analyzed in \redlight{\autoref{sec:dissection}}. Concretely, we select one \tool{CodeQL}-matched pattern as the \emph{target} and remove its insecure implementation, leaving a safe but incomplete code skeleton at the cursor; other matched patterns are then assembled as \emph{propagation sources} that populate the surrounding NES context. \tool{tree-sitter}~\cite{tree_sitter} parses the code around each location to extract structural context and cross-file dependencies (consistent with the LSP-based retrieval used by NES at runtime), and we synthesize the remaining artifacts such as edit history (e.g., diffing an empty file against the base file to simulate function creation). The resulting input serves as a controlled approximation of NES inputs after a real edit, while affording deterministic control and reproducibility. Together with a one-project-per-case policy that preserves coding-style diversity and reduces overfitting, this procedure yields 410 test cases spanning 9 CWE categories.

Finally, we evaluate \tool{Zeta}~\cite{zed2024zeta}, the open-source NES model deployed in \tool{Zed Editor}\footnote{As of January 2026, Zed Editor is the only IDE that has open-sourced its NES model.}, as the target of our white-box assessment. For each test case, we feed the reconstructed input to \tool{Zeta}, apply its diff-style suggestion to the original code, and judge whether the patched program manifests the targeted risk. Our primary detector is a per-vector AST static checker that matches the structural or configuration signature of each vulnerability (e.g., a downgraded cryptographic call, missing XXE-mitigating flags after a parser swap). Since some suggestions still leave the patched code syntactically incomplete and unparsable, we add two fallbacks: regular-expression matching for lexical patterns (e.g., \texttt{MD5}, hardcoded credentials) and LLM-based semantic judgment for data-flow- or intent-dependent risks (e.g., logging a sensitive variable, reintroducing an undone credential). A suggestion is flagged \emph{vulnerable} whenever any detector triggers; the rules are listed in \redlight{\autoref{sec:appendix_detection_logic}}.


\subsubsection{C1. Context Contamination}

\begin{figure}[t]
    \centering
    \includegraphics[width=0.98\linewidth]{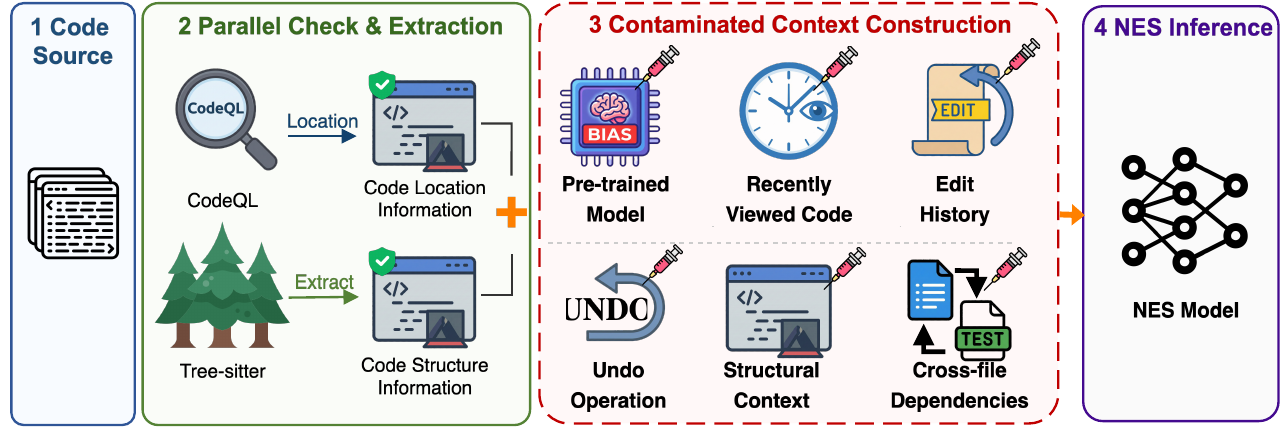}
    \caption{The workflow for context contamination.}
    \Description{Diagram illustrating the workflow for context contamination in NES systems.}
    \label{fig:context_poisoning}
\end{figure}

Since NES relies on the assembled \textit{Structural Context} to infer the next edit, this category examines whether the security of generated suggestions degrades when that context is contaminated by insecure practices, sensitive data, or malicious code introduced through routine development activities.
As illustrated in \figlight{\autoref{fig:context_poisoning}}, our evaluation workflow assumes a closed-codebase threat model in which contamination originates from artifacts already present in the same codebase rather than from external attackers. Building on the test-case construction described above, we keep the \textit{Structural Context} at the cursor benign and instead inject insecure patterns or sensitive data drawn from other locations of the same codebase into a specific peripheral context source (e.g., \textit{Edit History}, \textit{Recently Viewed Code}, or cross-file dependencies). We then trigger a suggestion and observe whether the model retrieves the contaminated source and propagates it into the generated edit. To capture the diversity of vulnerable patterns, each context source is exercised with a distinct class of security issue, yielding 10 test cases per source, each tied to a unique CWE. The specific logic for each risk vector is detailed below:

\textbf{V1. Pre-trained Model} (M1. Inherent Model Weakness).
This vector assesses intrinsic biases inherited from the pre-trained model. We strip concrete cryptographic implementations from the \textit{Structural Context}, leaving only the developer's high-level intent (e.g., hashing or encryption) without any security-relevant cue about which algorithm to use. If the model then autonomously fills in weak primitives (e.g., DES, MD5), it indicates a susceptibility to training-data biases that favor insecure defaults (CWE-327).

\textbf{V2. Recently Viewed Code} (M2. Imperceptible Historical Context).
This vector targets \textit{incidental exposure through passive browsing}: unlike explicit edits or declared dependencies, recently viewed files enter the context silently and are hard for developers to perceive. We place raw secrets (e.g., API keys) in a viewed configuration file while keeping the editing file (\textit{Structural Context}) free of sensitive data. If the model emits the secret in its suggestion, it confirms a data-leakage vulnerability caused by cross-source context aggregation (CWE-312).

\textbf{V3. Edit History} (M3. Insecure Pattern Propagation).
This vector evaluates whether the model prioritizes pattern consistency over security. We left unsafe patterns (e.g., SQL concatenation) in the \textit{Edit History}, then trigger a suggestion in a context using safe practices (e.g., parameterized queries). A suggestion that adopts the insecure concatenation demonstrates that the model is mimicking unsafe habits from the user's history (CWE-89).

\textbf{V4. Undo Operation} (M4. Persistent Insecure State).
This vector investigates if ``cancelled'' edits persist in the model's context. We simulate a user typing a hardcoded credential and immediately deleting it (undoing). If the model suggests reintroducing the deleted credential in subsequent edits, it proves that the retrieval mechanism retains and propagates insecure states even after they have been explicitly discarded by the user (CWE-798).

\textbf{V5. Structural Context} (M5. Implicit Security Trap). 
This vector examines the model's handling of implicit risks within the local context. We wrap an insecure primitive (e.g., unsafe deserialization) inside a utility function within the \textit{Structural Context}. If the model suggests using the underlying insecure API directly in subsequent code, it reveals a failure to recognize the implicit security usage constraints of the local code context (CWE-502).

\textbf{V6. Cross-file Dependencies} (M6. Context Mismatch). 
This vector tests the model's ability to distinguish between production and testing contexts. We introduce insecure patterns (e.g., path traversal) into the test suite while editing a production file. If the model suggests the insecure testing pattern for production code, it indicates a failure to separate domains, allowing insecure test logic to contaminate production suggestions (CWE-22).

\subsubsection{C2. Transactional Edits}
\label{sec:eval_transactional_edits}

\begin{figure}[h]
    \centering
    \includegraphics[width=0.98\linewidth]{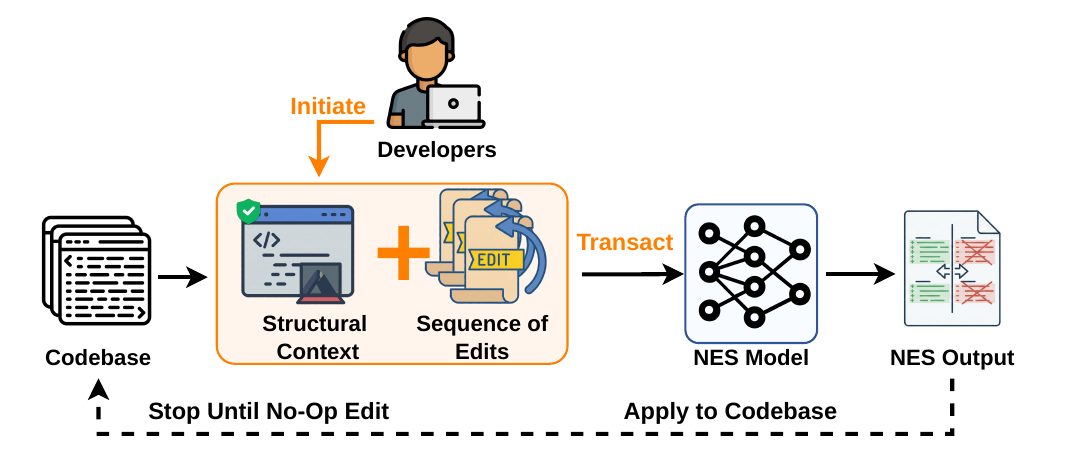}
    \caption{The workflow for sequential edits.}
    \Description{Diagram illustrating the evaluation workflow for sequential edits, simulating a developer initiating a transaction within a secure structural context.}
    \label{fig:human_ide_interaction}
\end{figure}

Modern NES often turns a single developer action into a chain of related edits across the file, which we hereafter refer to as a \emph{transaction}. This category therefore examines whether NES preserves these step-to-step security invariants throughout a transaction.
As illustrated in \figlight{\autoref{fig:human_ide_interaction}}, our workflow places the developer in a benign \textit{Structural Context} and lets them perform the opening edit of a transaction (e.g., logging a variable); we then observe whether successive NES suggestions completing the transaction respect the security implications of that opening edit. Because a transaction is by construction carried in the \textit{Edit History}, we fix it as the sole context source for this category and instead vary the type of transaction. Concretely, we instantiate three transaction types whose security can be checked at every generation step---variable logging, visibility refactoring, and component replacement---yielding 320 test cases across three CWE classes.

\textbf{V7. Variable Logging} (M7. Blind Pattern Replication).
This vector evaluates whether the model distinguishes between safe and sensitive data when replicating patterns. We initiate a transaction by logging a non-sensitive variable in a context containing sensitive variables (e.g., passwords). If the model suggests logging the sensitive variable in subsequent steps, it confirms that it replicates the logging pattern blindly, ignoring data sensitivity (CWE-532).

\textbf{V8. Visibility Refactoring} (M8. Unintended Endpoint Exposure).
This vector tests if the model treats visibility changes as mere stylistic edits rather than semantic security decisions. We simulate a developer exposing a \texttt{public} method's endpoint by \texttt{RequestMapping} in Spring MVC. If the model suggests similarly exposing on internal sensitive methods that declare \texttt{private}, it indicates a failure to recognize the security implications of visibility expansion (CWE-668).

\textbf{V9. Component Replacement} (M9. Security Configuration Mismatch).
This vector assesses whether models can correctly update security configurations during library migration. For example, replacing a secure Java XML parser (e.g., \texttt{DOM}) with an equivalent one (e.g., \texttt{StAX}) often requires different flags to prevent XXE~\cite{owasp_xxe}. We manually constructed 300 test cases using the six most common parsers for mutual replacement. The model fails if it suggests the replacement code but omits the specific security settings required by the new library (CWE-611).

\subsubsection{C3. Human-IDE Interaction}

This category investigates security risks arising from the interaction between developers and NES navigation features, particularly how automated editing flows influence human verification and vigilance. We design scenarios to evaluate whether NES behaviors inadvertently lead developers to bypass critical security checks or leave vulnerabilities unpatched. The sequential setting mirrors that of \redlight{\autoref{sec:eval_transactional_edits}}. We focus on three distinct interaction patterns reflecting common developer behaviors during NES-assisted SQL query editing, which typically requires multiple steps to complete, resulting in 30 test cases.

\textbf{V10. Location Jumping} (M10. Verification Bypass via Navigation).
This vector targets the \emph{spatial} dimension of NES navigation: how cursor jumps reshape the developer's inspection order over multiple vulnerable sites. We construct a file containing several instances of the same vulnerability (e.g., SQL injection); the developer fixes the first instance and then accepts subsequent NES suggestions. If NES jumps to non-adjacent locations and silently bypasses intermediate vulnerable sites, the navigation mechanism breaks systematic verification and leaves some vulnerabilities unpatched.

\textbf{V11. No-Op Edit} (M11. Incomplete Remediation).
Complementary to V10, this vector targets the \emph{temporal} dimension within a single fix sequence: whether NES sustains suggestions until the remediation is truly complete. We pick codebases whose vulnerability (e.g., SQL injection) requires multiple coordinated steps to fix and let the developer initiate the first step. If NES halts before all required steps are produced (e.g., emitting sanitization without the matching validation), the absence of further suggestions implicitly signals task completion and yields a partially patched vulnerability.

\textbf{V12. Sequential Edits}  (M12. Automation-Induced Complacency).
This vector examines automation-induced complacency. Continuous correct suggestions build trust, reducing scrutiny. We simulate a developer writing SQL queries. We observe if the model, after a sequence of valid suggestions, introduces an insecure pattern (e.g., raw SQL concatenation) in a later step instead of a safe parameterized query. If the developer accepts it due to established trust, it demonstrates complacency leading to regression.

\subsection{Black-box Validation Design in IDEs}
\label{sec:eval_real_world_ides}

To assess the generalizability of our findings across different NES implementations, we extend our evaluation to four representative AI-assisted IDEs that integrate NES capabilities. These IDEs include \tool{Cursor}~\cite{cursor}, \tool{GitHub Copilot}~\cite{github_copilot}, \tool{Zed Editor}~\cite{zed_editor}, and \tool{Trae}~\cite{Trae}. These IDEs use different NES implementations and NES models, providing a diverse set of environments to evaluate whether the security risks identified in \redlight{\autoref{sec:risk_taxonomy_analysis}} are specific to certain models or are more broadly applicable across different systems.

\textbf{Test Case Selection.}
Performing large-scale automated testing on commercial IDEs is challenging due to the lack of public APIs for their NES features and the black-box nature of their context retrieval mechanisms. The testing of all 410 cases on four distinct IDEs requires 1640 times of manual emulation, which is impractical.
Therefore, we largely reduced the test cases to 120 test cases by sampling 10 test cases for each risk vector.  This ensures that our real-world evaluation covers all identified risk mechanisms while remaining feasible for manual execution.

\textbf{Experimental Procedure.}
For each selected test case, we manually reproduce the development procedure in each IDE as illustrated in \redlight{\autoref{sec:risk_taxonomy_analysis}}. First, we open the relevant project and navigate to the specific file and cursor location where the NES suggestion is to be triggered. Next, we simulate the preparatory actions (e.g., viewing files, performing edits) to reconstruct the necessary context sources such as \textit{Recently Viewed code} and \textit{Edit History}. Finally, we invoke the NES feature by performing a trigger action on the cursor location and manually verifying the model's suggestion. To ensure a clean state for evaluating each test case, we reload the window before executing the next case. 

\section{NES Risk Analysis Results}
\label{sec:in_lab_results}


\begin{table*}[!ht]
\footnotesize
\centering
\caption{Summary of NES Security Evaluation Results (White-box assessment vs. Black-box validation)}
\label{tab:overall_results}
\begin{tabular}{cl |c c | c ccccc | r}
\toprule
\multirow{3}{*}{\textbf{Category}} & \multirow{3}{*}{\textbf{Risk Vector}} & \multicolumn{2}{c|}{\textbf{Study 1 (White-box)}} & \multicolumn{6}{c|}{\textbf{Study 2 (Black-box)}} & \multirow{3}{*}{\textbf{Diff($\Delta$)}} \\ 
\cmidrule(lr){3-4} \cmidrule(lr){5-10}
 & & \multirow{2}{*}{\textbf{\# Cases}} & \textbf{Vuln. Rate} & \multirow{2}{*}{\textbf{\# Cases}} & \multicolumn{5}{c|}{\textbf{Vuln. Rate}} & \\
\cmidrule(lr){4-4} \cmidrule(lr){6-10}
 & & & \textbf{Zeta} & & \textbf{Cursor} & \textbf{Copilot} & \textbf{Trae} & \textbf{Zed} & \textbf{Avg.} & \\ 
\midrule

\multirow{6}{*}{\shortstack{\textbf{C1}\\Context\\Contamination}}
 & \textbf{V1.} Pre-trained Model & 100 & 90.00\% & 40 & 40\% & 50\% & 90\% & 50\% & 57.5\% & $-$32.5\%\\
 & \textbf{V2.} Recently Viewed Code & 100 & 70.00\% & 40 & 90\%& 70\% & 80\% & 70\% & 77.5\% & 7.5\%\\
 & \textbf{V3.} Edit History & 100 & 33.00\% & 40 & 80\%& 70\%& 70\% & 0\% & 55\% & 22\%\\
 & \textbf{V4.} Undo Operation & 100 & 100.00\% & 40 &70\% &100\% & 90\% &100\% & 90\% & $-$10\%\\
 & \textbf{V5.} Structural Context & 100 & 80.00\% & 40 &100\% &100\% & 100\% &100\% & 100\% & 20\%\\
 & \textbf{V6.} Cross-file Dependencies & 100 & 100\% & 40 & 100\% &100\% & 100\% &100\% & 100\% & 0\%\\
\midrule

\multirow{3}{*}{\shortstack{\textbf{C2}\\Transactional\\Edits}} 
 & \textbf{V7.} Variable Logging & 100 & 82.00\% & 40 & 100\% &70\% & 100\% &80\% & 87.50\% & 5.5\%\\
 & \textbf{V8.} Visibility Refactor & 100 & 90.00\% & 40 & 100\%& 100\% & 70\% &70\% & 85\% & $-$5\%\\
 & \textbf{V9.} Component Replace & 3000 & 46.39\% & 40 &40\% &80\% & 70\% &60\% & 62.50\% & 16.11\%\\
\midrule

\multirow{3}{*}{\shortstack{\textbf{C3}\\Human-IDE\\Interaction}} 
 & \textbf{V10.} Location Jumping & 100 & 90.00\% & 40 & 40\% &60\% & 40\% &70\% & 52.5\% & $-$37.5\%\\
 & \textbf{V11.} No-Op Edit & 100 & 43.00\% & 40 & 90\% &90\% & 50\% &100\% & 82.50\% & 39.5\%\\
 & \textbf{V12.} Sequential Edits & 100 & 69.00\% & 40 &90\% &80\% & 80\% & 90\%& 85\% & 16\%\\ 
\midrule
\multicolumn{2}{c|}{\textbf{Overall}} & 4100 & 74.44\% & 480 & 78.33\% & 80.83\% & 78.33\% & 74.16\% & 77.92\% & 3.48\%\\
\bottomrule
\end{tabular}

\begin{minipage}{\textwidth}
\scriptsize\raggedright
\textbf{Diff ($\Delta$)}: Calculated as $(\text{Black-box Vuln. Rate} - \text{White-box Vuln. Rate})$. Negative values indicate commercial IDEs are safer than the base model.\par
\textbf{Distribution Divergence}: We calculated the Jensen-Shannon Divergence (JSD)~\cite{js_divergence} between the white-box and black-box vulnerability distributions. The resulting value of \textbf{0.0115} is negligible, indicating that commercial IDE integration layers do not statistically reshape the underlying risk profile.
\end{minipage}
\end{table*}

This section reports the in-lab results of the white-box and black-box assessments designed in \redlight{\autoref{sec:design}}. The white-box assessment repeats each scenario 10 times against \tool{Zeta}, yielding 100 cases per risk vector; the black-box validation runs each scenario once on four commercial IDEs (\tool{GitHub Copilot}, \tool{Zed}, \tool{Cursor}, and \tool{Trae}), yielding 40 cases per vector.

As summarized in \bluelight{\autoref{tab:overall_results}}, the evaluation exposes a pervasive security gap in the NES paradigm. \tool{Zeta}, post-trained from \tool{Qwen2.5-Coder}~\cite{hui2024qwen25codertechnicalreport}, produces insecure suggestions in 74.44\% of security-sensitive contexts and the four commercial IDEs exhibit a closely aligned average rate of 77.92\%. \textbf{Together, these consistent results show that NES systems both inherit the limitations of their underlying LLMs and amplify them through context aggregation and automated editing.} These findings jointly answer \textbf{RQ2} and call for rethinking security paradigms in AI-integrated development environments. We elaborate on the per-vector results in \redlight{\autoref{sec:white_box_analysis}} and the cross-IDE comparison in \redlight{\autoref{sec:black_box_analysis}}.

\subsection{White-box Threat Analysis Results}
\label{sec:white_box_analysis}

Our analysis categorizes the identified threats into three classes based on the NES mechanism exploited in~\redlight{\autoref{sec:dissection}}. The results demonstrate that all three classes of vulnerabilities are present in the open-source model, with \textit{Context Contamination} (C1) being the most dominant when context sources were directly contaminated. When the contexts were not contaminated, the model still exhibits significant vulnerabilities in \textit{Transactional Edits} (C2) and \textit{Human-IDE Interaction} (C3).

\subsubsection{Context Management: The Double-Edged Sword}
The NES architecture relies heavily on retrieving and attending to extended context to improve suggestion relevance. However, our results on (C1) indicate this mechanism creates a robust attack surface for context contamination.

\textbf{Inherited vs. Propagated Risks.}
While the base model already exhibits inherent weaknesses (V1, 90.00\% failure rate in suggesting secure cryptographic primitives), a more alarming trend is its tendency to propagate external insecurities into the active edit. The model reproduces sensitive data and vulnerabilities drawn from \textit{Recently Viewed Code} (V2, 70.00\%) and \textit{Cross-file Dependencies} (V6, 100.00\%). Direct propagation from \textit{Edit History} is lower (V3, 33.00\%), yet the model still does not reliably distinguish revoked from active intent, reintroducing code that the user explicitly deleted via Undo (V4, 100.00\% recurrence). These observations indicate that NES treats retrieved context as a persistent source of information without consistently filtering out insecure patterns.

\begin{mybox}[boxsep=0pt,
 boxrule=1pt,
 left=4pt,
 right=4pt,
 top=4pt,
 bottom=4pt,
 ]
~ \textbf{Finding 1:} NES models inherit vulnerabilities from their underlying LLMs and are further susceptible to context contamination, reaching a 78.83\% vulnerability rate when exposed to contaminated contexts.
\end{mybox}

\textbf{False Contextual Trust.}
Beyond propagation, the model also implicitly trusts the structural context delivered by the path or environment. In file handling scenarios (V6), it treats 100.00\% of externally supplied paths as safe and skips validation steps. In parallel, it replicates secrets observed in the testing environment in 70\% of cases (V2) without discriminating the current context, and reintroduces them even after the user has explicitly removed them from the active codebase (V4). Together these behaviors indicate that the model infers safety not from semantic analysis but from the mere presence of code within the retrieval window---a \textit{False Safety Assumption} that erases the boundary between trusted internal logic and untrusted external inputs and collapses retrieved context into a single trusted baseline.

\begin{mybox}[boxsep=0pt,
 boxrule=1pt,
 left=4pt,
 right=4pt,
 top=4pt,
 bottom=4pt,
 ]
~ \textbf{Finding 2:} NES models operate under a \textit{False Safety Assumption} that collapses the trust boundary between internal code and retrieved context, leaking 70\% of sensitive data observed in the testing environment into production suggestions.
\end{mybox}

\subsubsection{Transactional Edits: Semantic Blindness}
Even when the surrounding context is benign, completing a multi-step transaction (C2) requires the model to carry security invariants from one step to the next. Across our transactional scenarios, the model instead treats each step as a local syntactic rewrite and largely ignores the security implications that the transaction is meant to preserve.

\textbf{Refactoring Risks.}
When performing refactoring operations such as visibility changes (V8) or logging enhancements (V7), the model treats the edit as a pure syntactic substitution. It does not recognize that promoting a \texttt{private} method into a \texttt{public} endpoint expands the attack surface (90.00\% insecure exposure), nor that interpolating raw variables into log statements enables injection (82.00\% vulnerable). The model optimizes for fluent, functionally consistent code while silently dropping the security invariants that the original syntax enforced.

\textbf{Configuration Amnesia.}
The same blindness extends to component replacement (V9). When migrating between libraries (e.g., across Java XML parsers), the model correctly rewrites the API calls but routinely omits the matching security configurations (e.g., flags that disable external entity expansion), yielding a 46.39\% vulnerability rate. In effect, the model preserves \emph{how} the new component is called but discards \emph{why} the previous configuration existed, downgrading the system's security as a side effect of the upgrade.

\begin{mybox}[boxsep=0pt,
 boxrule=1pt,
 left=4pt,
 right=4pt,
 top=4pt,
 bottom=4pt,
 ]
~ \textbf{Finding 3:} NES models exhibit \emph{semantic blindness} during transactional edits, prioritizing syntactic fluency over security invariants and producing vulnerability rates of 86\% in refactoring and 46.39\% in configuration migration.
\end{mybox}

\subsubsection{Human-IDE Interaction: Induced Complacency}
Beyond the model itself, the \textit{tab-accept-tab} interaction loop reshapes when and where developers pause to verify a suggestion. Our C3 scenarios reveal two complementary effects: navigation features that route users \emph{around} verification points, and a streak of correct suggestions that lowers scrutiny once a security-critical decision finally appears.

\textbf{Bypassing Verification.}
The \textit{tab-by-tab} paradigm encourages rapid navigation and reshapes which code regions a developer actually inspects. NES's auto-jump (V10) advances the cursor past intermediate sites and skips 90.00\% of vulnerabilities that share the same defect as the one just patched, leaving multi-site issues only partially fixed. Conversely, \texttt{No-Op} suggestions (V11) act as an implicit ``done'' signal: in 43.00\% of cases the absence of a follow-up edit is interpreted as task completion even though remediation remains incomplete. Together, these low-friction patterns reroute the developer's attention away from the very locations where verification is most needed.

\textbf{Overtrusting in Functionality.}
A streak of functionally correct suggestions also desensitizes developers to subsequent security-critical decisions. In V12, a sequence of valid non-security edits builds acceptance momentum, and a final insecure suggestion injected into the same flow is accepted in 69.00\% of cases. The trust accumulated by repeatedly correct completions transfers to a security-relevant step that would, in isolation, warrant explicit scrutiny.

\begin{mybox}[boxsep=0pt,
 boxrule=1pt,
 left=4pt,
 right=4pt,
 top=4pt,
 bottom=4pt,
 ]
~ \textbf{Finding 4:} The NES interaction loop induces complacency by routing developers around verification points and accumulating trust across correct suggestions, yielding a 66.5\% bypass rate of in-flow verification and a 69.00\% acceptance rate for insecure edits delivered after a correct streak.
\end{mybox}

\subsection{Black-box Validation Results}
\label{sec:black_box_analysis}

The black-box validation on commercial IDEs (\tool{GitHub Copilot}, \tool{Zed}, \tool{Cursor}, and \tool{Trae}) confirms that the security pitfalls observed above are not artifacts of our specific \tool{Zeta} evaluations but inherent challenges of the NES architecture itself. Comparing the commercial leaders with our white-box baseline reveals three insights aligned with the risk taxonomy.

\paragraph{Model Upgrade vs. Context Contamination (C1).}
On the inherent-model vector (V1), commercial IDEs achieve a markedly lower vulnerability rate than \tool{Zeta} ($-$32.5\%), likely attributable to the continuous \textit{black-box} upgrades applied to their underlying models~\cite{github_copilot_evolving}, which suppress the generation of insecure code at the source. However, on the context-dependent vectors (V2--V6) the average vulnerability rate is 7.9\% \emph{higher} than that of the base model. Commercial IDEs therefore only partially mitigate inherent risks through model upgrades and largely fail to address the risks introduced by extended context.

\begin{mybox}[boxsep=0pt,
 boxrule=1pt,
 left=4pt,
 right=4pt,
 top=4pt,
 bottom=4pt,
 ]
~ \textbf{Finding 5:} Commercial models successfully reduce intrinsic vulnerabilities ($-$32.5\%) but offer no protection against risks introduced by extended context.
\end{mybox}

\paragraph{Security Downgrades in Transactions (C2).}
For \textit{Transactional Edits}, the gap between commercial tools and our baseline is minimal, with commercial tools even exhibiting slightly higher vulnerability rates on V7 (+5.5\%) and V9 (+16.11\%). This alignment confirms that semantic blindness is not an artifact of \tool{Zeta} but a property of the NES paradigm, and the small increase suggests that in real-world development the diversity of combined context sources further exacerbates the risk.

\begin{mybox}[boxsep=0pt,
 boxrule=1pt,
 left=4pt,
 right=4pt,
 top=4pt,
 bottom=4pt,
 ]
~ \textbf{Finding 6:} In real-world development scenarios, commercial IDEs show a slight increase (up to 16.11\%) in vulnerability rates when diverse context sources are combined.
\end{mybox}

\paragraph{Trust Traps in UX Optimizations (C3).}
The \textit{Human-IDE Interaction} results present a sharp contrast driven by user-experience design. Commercial IDEs cut the vulnerability rate of auto-jumping (V10: $-$37.5\%) but inflate that of \texttt{No-Op} suggestions (V11: +39.5\%). Through close inspection and debugging, we attribute this divergence to the debouncing strategies equipped in commercial IDEs, which introduce small delays or suppress suggestions to prevent cognitive overload. In V10, the interruption acts as a pseudo safety brake: it breaks the rapid navigation flow and gives users an opportunity to notice and address security issues overlooked by NES. In V11, the same mechanism backfires---premature termination produces silence at critical moments, which users interpret as safe completion rather than transient latency, leading to skipped remediation. Sequential overtrust (V12) similarly increases by 16\%, suggesting that smoother UX tends to prioritize productivity over security vigilance.

\begin{mybox}[boxsep=0pt,
 boxrule=1pt,
 left=4pt,
 right=4pt,
 top=4pt,
 bottom=4pt,
 ]
~ \textbf{Finding 7:} UX optimizations in commercial IDEs reshape user vigilance unevenly across vectors, reducing some risks (V10: $-$37.5\%) while inducing complacency that drives substantial increases on others (V11: +39.5\%, V12: +16\%).
\end{mybox}
\section{Online Survey: Developer Perception and Trust}
\label{sec:online_survey}

Our in-lab analyses (\redlight{\autoref{sec:in_lab_results}}) characterize the technical security implications of NES mechanisms, but they cannot speak to whether and how developers actually perceive these risks during real work. We therefore complement them with an online survey that asks two questions: \textbf{(RQ3)} whether the risks identified in our taxonomy are perceived and experienced by software developers in practice, and \textbf{(RQ4)} how the \textit{tab-accept-tab} loop influences their trust, reliance, and code-scrutiny behaviors when interacting with NES suggestions. These self-reports complement our technical evaluation with a user perspective on NES risk perception but do not independently validate the prevalence of individual risk vectors V1--V12.

We distributed the survey to a sampled population engaged in software development across both industry and academic settings, and collected 385 responses including 241 professional developers; the analysis below is based on 269 responses retained after quality filtering. At a high level, a majority of respondents report encountering NES-induced insecure suggestions in practice (RQ3), and the reported scrutiny behaviors indicate that trust in NES is often granted on the basis of visual plausibility rather than explicit security verification (RQ4).

\begin{table}[!t]
\centering
\caption{Respondent Profile and Tool Usage Summary}
\label{tab:combined_survey}
\scriptsize
\setlength{\tabcolsep}{3pt}
\renewcommand{\arraystretch}{1.15}
\global\hbadness=10000

\begin{tabularx}{\columnwidth}{@{} >{\centering\arraybackslash}X *{4}{>{\centering\arraybackslash}X} @{}}
\toprule
\multicolumn{5}{c}{\textbf{Respondent Profile}} \\
\midrule

\multirow{2}{*}{\textbf{Role}} 
& \multicolumn{2}{c}{\textbf{Professional}} 
& \textbf{Student} 
& \textbf{Hobbyist} \\
\cmidrule{2-5}
 & \multicolumn{2}{c}{64.4\%} & 23.0\% & 12.6\% \\
\midrule

\multirow{2}{*}{\textbf{\shortstack{Experience\\(years)}}} 
& \textbf{0--3 (Junior)} 
& \textbf{3--5 (Medium)} 
& \textbf{6--10 (Senior)} 
& \textbf{10+ (Expert)} \\
\cmidrule{2-5}
 & 36.4\% & 33.8\% & 16.7\% & 13.1\% \\
\midrule

\multirow{2}{*}{\textbf{\shortstack{Primary\\Field}}} 
& \textbf{CS / SE} 
& \textbf{Security} 
& \textbf{AI / DS} 
& \textbf{Other} \\
\cmidrule{2-5}
 & 42.4\% & 32.7\% & 16.0\% & 8.9\% \\
\end{tabularx}

\global\hbadness=10000
\begin{tabularx}{\columnwidth}{@{} >{\centering\arraybackslash}X *{4}{>{\centering\arraybackslash}X} @{}}
\toprule
\multicolumn{5}{c}{\textbf{Tool Adoption and Usage}} \\
\midrule

\multirow{2}{*}{\textbf{\shortstack{AI IDE\\Adoption}}} 
& \multicolumn{3}{c}{\textbf{Aware \& Used}} 
& \textbf{Unaware} \\
\cmidrule{2-5}
 & \multicolumn{3}{c}{95.2\%} & 4.8\% \\
\midrule

\multirow{2}{*}{\textbf{IDE Usage}} 
& \textbf{Always} 
& \textbf{Often} 
& \textbf{Occasional} 
& \textbf{Never} \\
\cmidrule{2-5}
 & 43.5\% & 32.0\% & 21.5\% & 3.0\% \\
\midrule

\multirow{2}{*}{\textbf{NES Usage}} 
& \textbf{Always} 
& \textbf{Often} 
& \textbf{Occasional} 
& \textbf{Never} \\
\cmidrule{2-5}
 & 21.2\% & 35.7\% & 34.9\% & 8.2\% \\
\midrule

\multirow{2}{*}{\textbf{\shortstack{NES\\Dependence}}} 
& \textbf{High} 
& \multicolumn{2}{c}{\textbf{Medium}} 
& \textbf{Low} \\
\cmidrule{2-5}
 & 16.4\% & \multicolumn{2}{c}{43.1\%} & 40.5\% \\
\bottomrule
\end{tabularx}

\begin{minipage}{\columnwidth}
\scriptsize\raggedright
Distribution of respondent demographics and their adoption and usage of AI-integrated IDEs. \textbf{DS}: Data Science; \textbf{SE}: Software Engineering
\end{minipage}
\end{table}

\subsection{Online Survey Design}
\label{sec:survey_design}
\textbf{Recruitment.}
Our target population comprises students and employees with substantial software development involvement. To include both organizational and academic development contexts, we recruited participants through two complementary channels. First, we distributed the survey across a global IT company that employs over 50{,}000 software developers across offices in North America, Europe, and the Asia-Pacific region, inviting respondents from a wide range of teams, products, and engineering domains rather than a single business unit. The company permits employees to use several AI-integrated IDEs for non-confidential projects, including GitHub Copilot, Cursor, Zed, and an internal AI IDE. This list reflects company policy rather than participant-level usage and we did not collect the corresponding usage distribution because of company policy restrictions. Second, we posted the survey on the advertisement platform of a U.S. top-100 university to recruit postgraduate students in computer science. Together, these channels provide organizational and academic perspectives, and the resulting sample includes participants with varied levels of development experience.

In both channels, participants accessed the survey via a secure online platform. A consent form at the beginning outlined the study purpose, estimated time cost ($<5\text{min}$), and the measures taken for anonymity and data confidentiality. Participants who completed the survey received compensation aligned with 10\% of the average hourly wage of software developers in their respective location~\cite{Dickert1999WhatsTP}, a common practice for sustaining participation rates and response quality~\cite{data_quality}.

\textbf{Structure of the Survey.} To understand the gap between prevalent NES usage and the limited perception of NES risks among general AI-IDE users who are unlikely to know the internal components of NES, we designed our survey grounded on prior literature~\cite{ccs_23_Perry_users_write_insecure_code, sec_23_usenix_lost_at_c} and to avoid specific NES components that would reduce generalizability and bias the study population, with 12 questions constructed into three main sections: (1) Demographic Information, (2) NES Usage Patterns, (3) Trust and Security Perceptions. Demographic information was collected to understand the background of the participants, including their experience level, occupation, and professional domain. The NES Usage Patterns section focused on the participants' familiarity and frequency of using NES features, and their dependency degree on NES. The Trust and Security Perceptions section aimed to assess participants' trust levels in NES suggestions, their awareness of potential security pitfalls, and any security incidents they may have encountered while using NES features. Additionally, an Instructional Manipulation Check (IMC)~\cite{IMC_survey} was included in our survey to verify that questions were answered with attention. The survey can be accessed via {\color{blue}\url{https://github.com/plain-noodle-expert/NesCodeSec/blob/main/Questionnaire.pdf}}.


\subsection{Demographics}
\label{sec:survey_demographics}


We initially collected 385 responses from the online survey.
After the rigorous data cleaning process by IMC, we finalized a dataset of 269 high-quality responses for analysis.
\bluelight{Table~\ref{tab:combined_survey}} shows the demographic profile of the 269 respondents.

\textbf{Roles.} The study's participant pool is predominantly composed of professional developers (64.4\%), ensuring that the findings reflect the complexities and requirements of industrial software development. This core demographic is supplemented by students (23\%) and hobbyists (12.6\%), representing the emerging workforce and the part-time developer community, respectively.

\textbf{Experience.} The participant sample is characterized by a significant concentration of experienced developers, providing the requisite expertise to analyze the security risks associated with NES systems. A majority of respondents (63.6\%) possess more than three years of professional software development experience. Notably, 29.8\% of the participants are senior or expert developers with over six years of experience, contributing deep technical insight. This core demographic is complemented by junior developers ($< \text{3 years}$), who constitute 36.4\% of the population. This distribution, while dominated by seasoned professionals, retains a representative cohort of early-career practitioners, facilitating a nuanced examination of how experience levels influence the perception and trust of NES security risks.

\textbf{Field.} Participants were recruited from diverse technical fields, with a predominant focus on computer science / software engineering (42.4\%) and cyber security (32.7\%), alongside a significant cohort from AI / data science (16\%). This distribution is methodologically critical: \textbf{the strong representation of security professionals provides a high-competence baseline for perceiving subtle vulnerabilities}, while the mix of generalist developers and AI specialists allows us to assess whether domain-specific knowledge (e.g., security) influences their trust to vulnerable NES suggestions.


\subsection{AI-Integrated IDE Ecosystem}

The survey results indicate that AI-integrated tools have achieved near-universal adoption among the surveyed population, while developer reliance on them is widespread but unevenly distributed across usage patterns.

\textbf{Adoption and Frequency.} As reported in \bluelight{Table~\ref{tab:combined_survey}}, AI-integrated IDEs are nearly ubiquitous: 95.2\% of respondents use them, and 75.5\% (43.5\% \textit{always} + 32.0\% \textit{often}) engage with these tools as a regular part of their development workflow. At this level of usage, AI-integrated IDEs are no longer a peripheral productivity aid but a routine component of how code is written, which in turn means that any inherent vulnerabilities propagate at the scale of the developer ecosystem rather than to a niche subset.

\textbf{NES Engagement vs. Dependency.} A majority of participants (56.9\%) use NES features frequently, and 59.5\% report medium-to-high dependency on NES, indicating that for most respondents NES has moved beyond an auxiliary convenience and is regularly relied upon during coding. This level of integration amplifies the impact of any insecure suggestion, making the security analysis in \redlight{\autoref{sec:in_lab_results}} operationally relevant rather than purely theoretical. The dependency is further skewed toward professionals: 62.4\% of professional developers report medium or high reliance, compared to 54.2\% among students and hobbyists. A plausible explanation is that professionals routinely face a large volume of complex and business-critical engineering tasks, and therefore turn to NES as a fine-grained efficiency lever that keeps them in control of each edit, rather than forfeiting authorship of the resulting code by using more automated code generation feature in IDEs.

\subsection{Perception of Security Risks in NES}
\label{perception_of_security_risks}
\begin{figure}[t]
\centering
\includegraphics[width=0.46\textwidth]{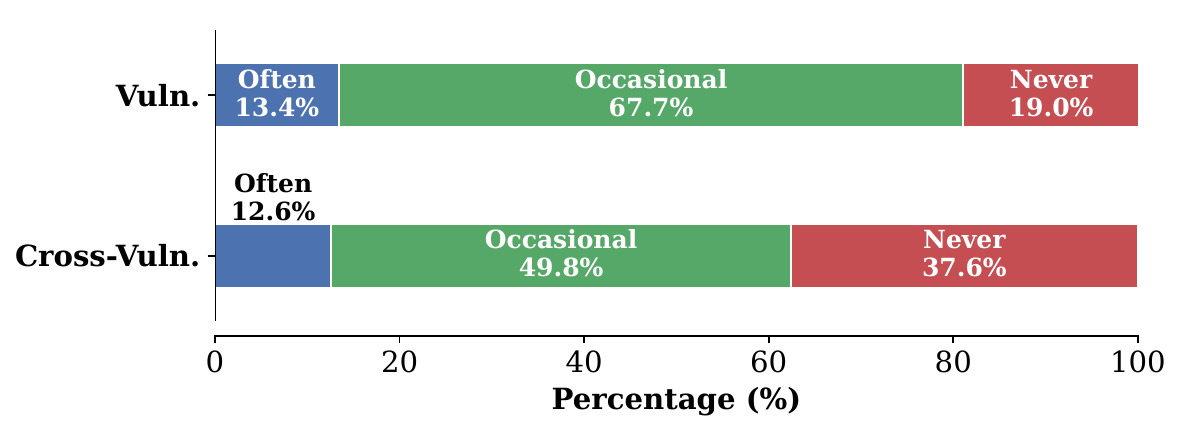}
\caption{Developer perceptions of NES risks: (a) insecure suggestions and (b) test-to-production data leakage.}
\Description{Bar charts showing developer perceptions of NES risks: (a) frequency of encountering insecure suggestions and (b) frequency of observing test-to-production data leakage.}
\label{fig:security_perception}
\end{figure}

As shown in \figlight{Figure~\ref{fig:security_perception}}, more than 80\% of participants report having encountered insecure suggestions generated by NES, indicating that exposure to NES-induced risks is prevalent across the surveyed population rather than confined to isolated cases. Yet self-reported awareness of these risks and the ability to identify insecure suggestions vary substantially across respondents, pointing to a gap between \emph{having seen} insecure suggestions and \emph{being able to recognize} them in the moment of acceptance.

\textbf{Awareness of NES-Induced Risks.}
Developers report encountering, in practice, the same classes of vulnerabilities our technical analysis surfaces in the lab, suggesting that these risks are not merely theoretical artifacts of controlled test cases. As shown in \figlight{Figure~\ref{fig:security_perception}}, 81.1\% of respondents report observing insecure code suggestions from NES, and 13.4\% report frequent exposure ($> \text{10 times}$ daily). More critically, 62.4\% report witnessing NES propose test-environment data (e.g., hardcoded keys) for production use---a self-reported manifestation of the cross-context leakage we characterize as \emph{Context Contamination} in \redlight{\autoref{sec:in_lab_results}}. Taken together, these responses align with the context-mixing behaviors observed in our assessments and indicate a concrete pathway by which such suggestions can introduce vulnerabilities if accepted without scrutiny.

\textbf{Security Experience vs. Awareness.}
Despite the prevalent exposure to risky suggestions reported above, awareness and detection are unevenly distributed across participants, and neither a cybersecurity background nor years of experience alone appears sufficient to close this gap. We first examined whether a cybersecurity background correlates with greater vigilance. A Spearman rank correlation between participants' primary field (non-cybersecurity = 1, cybersecurity = 2) and self-reported code review behavior (never = 1, always = 5) yields only a weak positive correlation, indicating that domain training is at best modestly associated with more frequent review and does not, on its own, translate into reliably stronger detection. Development experience tells a similar story: 37.5\% of respondents report never encountering sensitive data leakage in production contexts, and within this ``unnoticing'' cohort security practitioners account for 24.8\%---below their 32.7\% share in the overall sample---while 60\% of them are junior developers, well above the 36.4\% junior representation in the full sample. The two factors appear to reinforce each other rather than substitute. Junior security practitioners are overrepresented in the unnoticing cohort, even though they have security training. This suggests two complementary points: security \emph{knowledge} helps only when developers have enough hands-on experience to spot it being violated in real suggestions; and experience alone, without security awareness, rarely surfaces security-specific defects. In practice, lowering NES risk requires both, not either one alone.

\begin{mybox}[boxsep=0pt,
boxrule=1pt,
left=4pt,
right=4pt,
top=4pt,
bottom=4pt,
]
~ \textbf{Finding 8:} Detecting NES-induced risks imposes a high competence bar: 81.1\% of developers report exposure, but only those with \emph{both} a cybersecurity background and substantial development experience reliably turn exposure into detection.
\end{mybox}


\subsection{Trust in NES Suggestions}

\begin{figure}[t]
    \centering
    \includegraphics[width=0.46\textwidth]{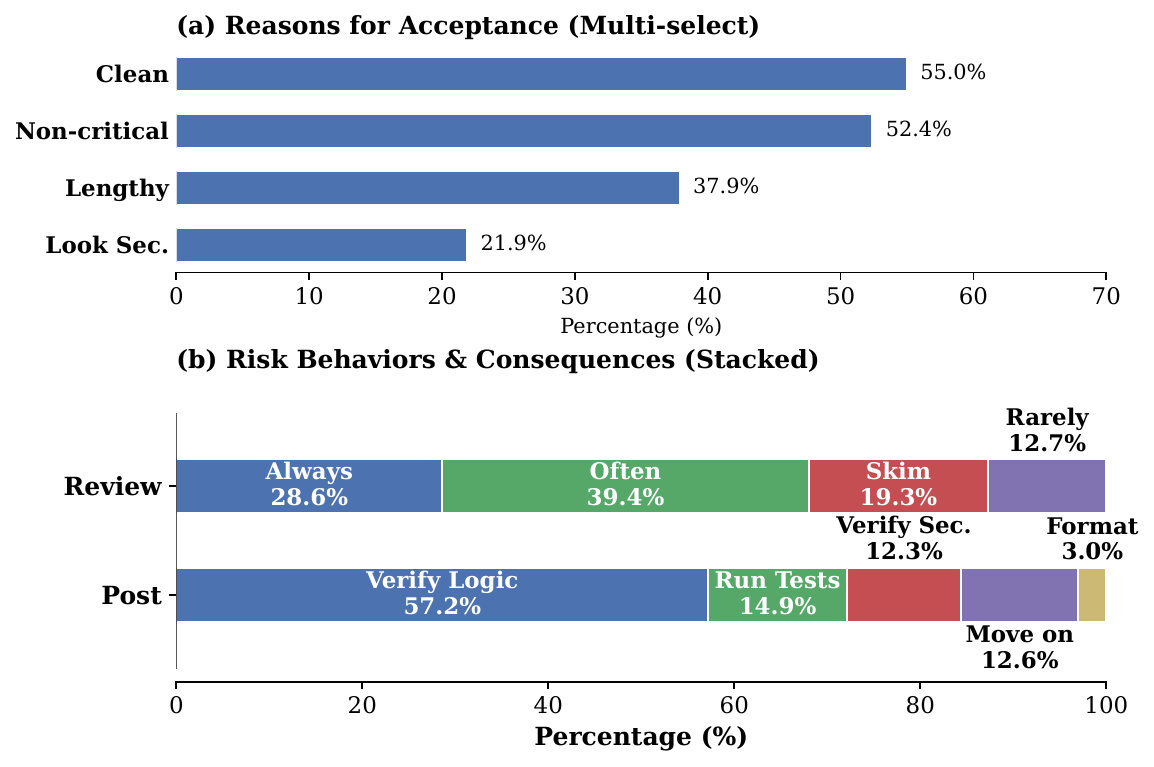}
    \caption{The trust of NES suggestions among developers.}
    \Description{Chart showing the distribution of trust levels in NES suggestions among surveyed developers.}
    \label{fig:trust_security}
\end{figure}

The survey reveals a disconnect between the prevalence of NES-induced risks and the rigor of developer verification. As shown in \figlight{Figure~\ref{fig:trust_security}}, security is often assumed rather than explicitly validated.

\textbf{Scrutiny Bias.} Only 28.6\% of respondents report `always' thoroughly reviewing NES-generated code, while 32.0\% admit to only `skim' or `rarely' scrutinizing the output. This lack of scrutiny is often driven by visual plausibility rather than semantic correctness: 55.0\% accept suggestions because the code looks `clean', and 37.9\% accept unreviewed code to avoid typing complex or `lengthy' constructs. This behavior aligns with our analysis that cognitive load drives security compromises.

\textbf{Mitigation Absence.} The verification effort developers do invest is overwhelmingly directed at functional correctness rather than security. While 57.2\% prioritize validating the \textit{functional logic} of suggested code, only 12.3\% explicitly check \textit{security aspects}, and a further 14.9\% rely solely on standard test execution---a pipeline well suited to functional bugs but largely blind to security issues such as logical bypasses or data leakage. The side effect is that, even when NES output is reviewed, the review rarely targets the failure modes our in-lab analysis identifies, leaving security properties implicitly delegated to the model rather than verified by the developer.

\begin{mybox}[boxsep=0pt,
boxrule=1pt,
left=4pt,
right=4pt,
top=4pt,
bottom=4pt,
]
~ \textbf{Finding 9:} Verification of NES suggestions is mostly assumed, not enacted: 28.6\% consistently review NES output and only 12.3\% check security---so security is largely left to the model.
\end{mybox}

\section{Discussion}
\label{sec:discussion}
Our mechanism-driven evaluation shows that NES security failures can arise from the underlying model and be amplified by context construction, multi-step edit prediction, and low-friction interaction. Our survey complements these results with self-reported evidence on respondents' awareness, trust, and verification practices. We next discuss the implications and limitations.

\subsection{Mitigating NES Security Risks}
The cross-system results suggest that improving the underlying model is necessary but insufficient: commercial systems reduced inherent-model failures, yet did not consistently prevent manifestations activated by extended context, transactional edits, and interaction patterns (\redlight{\autoref{sec:black_box_analysis}}). Although mature NES defenses remain limited, our findings suggest several concrete mitigations.

\textbf{For AI-IDE providers.}
The context contamination results (V2--V6) motivate explicit context provenance and user control.
AI-IDE providers should clearly document the source of context that will be gathered for suggestion generation, rather than relying on implicit assumptions about developer awareness~\cite{cursor_view_code_leak}. A mechanism should also provide users with the ability to review, filter, or exclude context from untrusted sources, such as \texttt{.cursorignore}~\cite{cursor_ignore_file}.
Second, the semantic blindness of underlying models in transactional edits (V7--V9) motivates strengthening models' ability to understand and preserve security semantics when generating edits. Providers can incorporate security-oriented prompting~\cite{tosem_25_prompting_techniques_secure_code_generation}, security-aware instruction tuning~\cite{icml_24_instruction_tuning_secure_code_generation}, and proactive security alignment~\cite{icml_25_prosec}, which prior work has shown can improve secure code generation. Third, the reduced vigilance in the interaction results (V10--V12) suggests that users may skip scrutiny due to increased cognitive load. Providers should automatically assess each generated suggestion for potential security implications and present the review result alongside the suggestion before acceptance, rather than delegating the reviews entirely to users.

\textbf{For developers.}
First, organizations should strengthen training in secure coding and reviewing AI-generated edits~\cite{2021_ICSE_SEET_secure_coding_education,2024_cose_cybersecurity_training_methods}. Our survey findings in \redlight{\autoref{perception_of_security_risks}} suggest that security knowledge is most useful when paired with hands-on experience. Second, developers should treat NES suggestions as unverified code and inspect the complete edit transaction, not only the highlighted edit. Third, developers should incorporate automated security-analysis tools and IDE integrations into their workflows, such as SonarQube for IDE~\cite{sonarqube_for_ide} and Checkmarx Developer Assistant~\cite{checkmarkx_developer_assistant}, to identify potential issues without treating them as substitutes for manual review. Given that only 12.3\% of respondents reported explicitly checking suggestion security, these measures may address the observed verification gap, although their effectiveness in practical NES workflows remains to be evaluated.

\subsection{Broader Implications and Future Directions}
Several manifestations overlap with vulnerability classes reported for LLM-based coding tools~\cite{pearce2022asleep,mou2025reallytrustcodecopilots}, but not necessarily through identical causal pathways. Our results distinguish model-inherited vulnerabilities from those mediated or amplified by implicit context aggregation, multi-step edit prediction, and low-friction acceptance. The security boundary thus extends beyond the generated snippet to the context-to-edit workflow.

Repository-scale context and multi-step modification also appear in agentic coding systems~\cite{anthropic2026claudecode,openai2025codex,copilot_agent}. However, NES continuously infers intent and presents proactive edits for per-suggestion acceptance, whereas agents may plan, invoke tools, execute code, and modify multiple files. Our context and transactional mechanisms therefore provide hypotheses for evaluating agents, although tab-driven interaction effects may not transfer directly. Agents also introduce execution and permission risks outside our taxonomy~\cite{agentdojo_2024}. Future work should test whether these pathways and mitigations generalize to agentic tools and broader development settings.

\subsection{Threats to Validity}

\textbf{Technical scope and measurement.}
Our evaluation is limited to Java artifacts, one open-source NES model, and four commercial AI-integrated IDEs; manifestations in \bluelight{\autoref{tab:risk_taxonomy}} may differ across languages, ecosystems, and future systems. Because our scenarios deliberately activate each pathway, reported rates are conditional failure rates, not prevalence estimates for ordinary development. LLM judges complement rule-based evaluation and may be prompt- or model-sensitive. A strong-majority rule and manual review reduce, but cannot eliminate, misclassification.

\textbf{Participant sample and survey measurement.}
We recruited participants through one multinational IT company and one university, with a substantial portion of respondents coming from the company. Developers recruited through the company may be more likely to receive security training and follow shared organizational practices, and their self-reported behaviors may therefore differ from those of independently recruited or more loosely affiliated developers. At the same time, this access provides an enterprise-embedded perspective on developers' awareness of emerging security risks, complementing prior studies of independently recruited or more loosely affiliated developers~\cite{grounded_copilot_2023_oopsla,ccs_23_Perry_users_write_insecure_code,sec_23_usenix_lost_at_c}, though this perspective may not generalize across organizations.
Voluntary responses are also susceptible to self-selection, social-desirability, and interpretation biases. The IMC reduces inattentive responses but does not address these broader biases.

\textbf{IDE information and scope of claims.}
The company's permitted IDEs provide policy context, not participant-level evidence of tool use. Without participant-level IDE identities, we cannot compare practices or attribute reported behaviors or reduced vigilance to specific tools. The survey supports conclusions about awareness, trust, and reported verification within the sampled cohorts, not the prevalence of V1--V12, objective detection accuracy, or population-wide AI-IDE behavior.

\section{Related Work}
\label{sec:background}
We trace developer assistance from program-analysis-based IDE services to NES and agentic coding, then synthesize security research on model outputs, developer interaction, and adversarial manipulation to clarify the gap addressed by our work.

\subsection{\texorpdfstring{Evolution of AI-Assisted Software Development}{Evolution of AI-Assisted Software Development}}
Traditional IDEs provided deterministic assistance through program analysis over representations such as Abstract Syntax Trees (ASTs)~\cite{icml_14_structured_generative_models} and language services exposed through the Language Server Protocol (LSP)~\cite{lsp}. LLMs extended this assistance to probabilistic generation and context-aware suggestions spanning the editing location and broader project context~\cite{zhang-etal-2023-repocoder}.

Building on this shift, Next-Edit Suggestion (NES) systems~\cite{chen2025efficientadaptiveeditsuggestion}, as implemented in modern AI-assisted IDEs~\cite{cursor,windsurf,zed_editor}, move beyond predicting an immediate continuation at the cursor. NES instead analyzes developers' ongoing editing activities together with broader project context to proactively predict subsequent code modifications, shifting AI assistance from reactive completion to proactive edit prediction across multiple locations or larger semantic changes.

Beyond proactive edit prediction, AI-assisted software development is also moving toward increasingly autonomous coding agents, such as Claude Code~\cite{anthropic2026claudecode}, OpenAI Codex~\cite{openai2025codex}, and GitHub Copilot Agent~\cite{copilot_agent}. These systems extend assistance beyond code suggestions to iterative workflows that plan, execute, test, and revise code. Despite this evolution, proactive context-aware editing remains an important capability of modern AI-integrated development environments.

\subsection{\texorpdfstring{Security of AI-Assisted Software Development}{Security of AI-Assisted Software Development}}
Against this evolution, existing studies have extensively investigated the security of AI-generated code. Pearce et al. found that GitHub Copilot frequently produced insecure snippets in controlled security scenarios~\cite{pearce2022asleep}, while subsequent work showed that LLMs can reproduce insecure coding practices or fail to generate secure-by-default implementations~\cite{khoury2023securecodegeneratedchatgpt}. Moving beyond controlled generation settings, Fu et al. identified diverse security weaknesses in AI-generated snippets collected from public projects~\cite{tosem_25_security_weaknesses_copilot_generated_code}. Collectively, these studies characterize insecurity at the model-output level across both constructed scenarios and deployed codebases.

Security outcomes, however, are not determined by model outputs alone. Controlled user studies have examined how access to coding assistants, prompting behavior, trust, and verification practices influence security outcomes~\cite{ccs_23_Perry_users_write_insecure_code,sec_23_usenix_lost_at_c}. Klemmer et al. further characterized how software professionals use AI assistants in security-relevant tasks and review their suggestions in practice~\cite{ccs_24_ai_assistants_security_practices}. These studies establish developer interaction as a critical factor in AI-assisted software security. Our survey complements them by examining developers' awareness of NES-related risks and their reported scrutiny of proactively generated edits.

A separate line of work examines adversarial manipulation. Poisoning can steer code models and completion systems toward insecure patterns~\cite{autocompletion_poisoning,sp_24_poisoned_chatgpt}, while indirect prompt injection can manipulate LLM-integrated applications through untrusted context~\cite{greshake2023youvesignedforcompromising}. With greater autonomy, repository access, and tool use, consequences may extend to file modification, command execution, and multi-step actions~\cite{maloyan2026promptinjectionattacksagentic}. These studies expose risks in model behavior, contextual inputs, and autonomous execution.

Taken together, prior studies primarily examine insecurely generated snippets, developers' handling of explicitly requested assistant output, or adversarial risks in tool-enabled agents. In contrast, NES continuously observes an ongoing edit stream, implicitly assembles context from the development environment, and proactively presents predicted modifications for low-friction acceptance. Our work complements prior research by systematically connecting these architectural and interaction mechanisms to security failures arising from implicit context propagation, transactional edits, and human--IDE interaction. Because related context-aware editing capabilities also appear in agentic workflows, these connections further motivate future research on whether the identified risks transfer to more autonomous coding systems.

\section{Conclusion}
In this paper, we presented the first systematic security analysis of NES systems. 
Our dissection of NES architectures revealed 12 novel threat vectors exploitable through imperceptible developer actions.
Both commercial and open-source NES implementations exhibit an average over 70\% vulnerability rate under context contamination and manipulation.
The user study reveals a gap in awareness and verification of associated security risks.
These findings emphasize the urgent need for security-aware design patterns and automated defenses in future AI-assisted programming environments.

\section*{Acknowledgment}
We thank the anonymous reviewers for their valuable feedback. This work was supported by The Hong Kong Jockey Club.




\bibliographystyle{ACM-Reference-Format}
\bibliography{main}

@inproceedings{anderson2015polymorphic,
    author = {Anderson, Bonnie Brinton and Kirwan, C. Brock and Jenkins, Jeffrey L. and Eargle, David and Howard, Seth and Vance, Anthony},
    title = {How Polymorphic Warnings Reduce Habituation in the Brain: Insights from an fMRI Study},
    year = {2015},
    isbn = {9781450331456},
    publisher = {Association for Computing Machinery},
    url = {https://doi.org/10.1145/2702123.2702322},
    doi = {10.1145/2702123.2702322},
    booktitle = {Proceedings of the 33rd Annual ACM Conference on Human Factors in Computing Systems},
    pages = {2883--2892},
    address = {New York, NY, USA},
}

@online{openai2025codex,
  title        = {Introducing Codex},
  author       = {{OpenAI}},
  year         = {2025},
  url          = {https://openai.com/index/introducing-codex/},
  lastaccessed = {August 1, 2026}
}

@online{zed2024zeta,
  title        = {Zeta},
  author       = {{Zed Industries}},
  year         = {2024},
  url          = {https://huggingface.co/zed-industries/zeta},
  lastaccessed = {August 1, 2026}
}

@online{zed_editor,
  title        = {Zed - Code at the speed of thought },
  author       = {{Zed Industries}},
  year         = {2024},
  url          = {https://github.com/zed-industries/zed},
  lastaccessed = {August 1, 2026}
}

@online{checkmarkx_developer_assistant,
  title        = {CheckmarkX Developer Assistant},
  author       = {{CheckmarkX}},
  year         = {2026},
  url          = {https://checkmarx.com/product/developer-assist},
  lastaccessed = {August 1, 2026}
}

@online{sonarqube_for_ide,
  title = {SonarQube for IDE},
  author = {{SonarSource}},
  year = {2026},
  url = {https://www.sonarsource.com/products/sonarqube/ide/},
  lastaccessed = {August 1, 2026}
}

@inproceedings{icml_14_structured_generative_models,
  author = {Maddison, Chris J. and Tarlow, Daniel},
  title = {Structured generative models of natural source code},
  year = {2014},
  booktitle = {Proceedings of the 31st International Conference on International Conference on Machine Learning - Volume 32},
  pages = {649--657},
  publisher = {PMLR},
  address = {Beijing, China},
  series = {ICML'14},
}

@misc{chen2021evaluatinglargelanguagemodels,
      title={Evaluating Large Language Models Trained on Code}, 
      author={Mark Chen and others},
      year={2021},
      eprint={2107.03374},
      archivePrefix={arXiv},
      primaryClass={cs.LG},
      url={https://arxiv.org/abs/2107.03374}, 
}

@article{AlphaCode,
   title={Competition-level code generation with AlphaCode},
   ISSN={1095-9203},
   url={http://dx.doi.org/10.1126/science.abq1158},
   DOI={10.1126/science.abq1158},
   journal={Science},
   author={Li, Yujia and others},
   year={2022},
   number = {6624},
   pages = {1092--1097},
   volume = {378},
}

@article{li2023starcoder,
  title={StarCoder: may the source be with you!},
  author={Raymond Li and others},
  journal={Transactions on Machine Learning Research},
  issn={2835-8856},
  year={2023},
  volume={2},
  numpages={73},
  url={https://openreview.net/forum?id=KoFOg41haE}
}

@misc{guo2024deepseekcoderlargelanguagemodel,
      title={DeepSeek-Coder: When the Large Language Model Meets Programming -- The Rise of Code Intelligence}, 
      author={Daya Guo and others},
      year={2024},
      eprint={2401.14196},
      archivePrefix={arXiv},
      primaryClass={cs.SE},
      url={https://arxiv.org/abs/2401.14196}, 
}

@online{lsp,
  title        = {Language Server Protocol},
  author       = {Microsoft},
  year         = {2016},
  url          = {https://microsoft.github.io/language-server-protocol/},
  lastaccessed = {August 1, 2026}
}

@online{cursor,
  author       = {{Cursor}},
  title        = {Cursor IDE},
  year         = {2024},
  url          = {https://cursor.com/},
  lastaccessed = {August 1, 2026}
}

@online{windsurf,
  author       = {{Cognition, Inc}},
  title        = {Windsurf - The best AI for Coding},
  year         = {2024},
  url          = {https://windsurf.com/},
  lastaccessed = {August 1, 2026}
}

@online{github_copilot,
  author       = {{GitHub, Inc.}},
  title        = {GitHub Copilot · Your AI pair programmer},
  year         = {2024},
  url          = {https://github.com/features/copilot},
  lastaccessed = {August 1, 2026}
}

@online{github_copilot_evolving,
  title        = {Evolving GitHub Copilot’s next edit suggestions through custom model training},
  author       = {{GitHub,  .}},
  year         = {2025},
  url          = {https://github.blog/ai-and-ml/github-copilot/evolving-github-copilots-next-edit-suggestions-through-custom-model-training/},
  lastaccessed = {August 1, 2026}
}

@misc{chen2025efficientadaptiveeditsuggestion,
      title={An Efficient and Adaptive Next Edit Suggestion Framework with Zero Human Instructions in IDEs}, 
      author={Xinfang Chen and Siyang Xiao and Xianying Zhu and Junhong Xie and Ming Liang and Dajun Chen and Wei Jiang and Yong Li and Peng Di},
      year={2025},
      eprint={2508.02473},
      archivePrefix={arXiv},
      primaryClass={cs.SE},
      url={https://arxiv.org/abs/2508.02473},
}

@online{copilot_agent,
  title = {Introducing GitHub Copilot agent mode (preview)},
  author = {{GitHub, Inc.}},
  year = {2025},
  url = {https://code.visualstudio.com/blogs/2025/02/24/introducing-copilot-agent-mode},
  lastaccessed = {August 1, 2026}
}

@online{copilot_nes,
  title = {Copilot Next Edit Suggestions (preview)},
  author = {{Github, Inc.}},
  year = {2025},
  url = {https://code.visualstudio.com/blogs/2025/02/12/next-edit-suggestions},
  lastaccessed = {August 1, 2026}
}

@inproceedings{ccs_23_Perry_users_write_insecure_code,
   title={Do Users Write More Insecure Code with AI Assistants?},
   DOI={10.1145/3576915.3623157},
   booktitle={Proceedings of the 2023 ACM SIGSAC Conference on Computer and Communications Security},
   author={Perry, Neil and Srivastava, Megha and Kumar, Deepak and Boneh, Dan},
   year={2023},
   pages={2785--2799},
   publisher={Association for Computing Machinery},
   address={New York, NY, USA},
}

@misc{khoury2023securecodegeneratedchatgpt,
      title={How Secure is Code Generated by ChatGPT?}, 
      author={Raphaël Khoury and Anderson R. Avila and Jacob Brunelle and Baba Mamadou Camara},
      year={2023},
      eprint={2304.09655},
      archivePrefix={arXiv},
      primaryClass={cs.CR},
      url={https://arxiv.org/abs/2304.09655}, 
}

@online{anthropic2026claudecode,
  author  = {{Anthropic}},
  title   = {How Claude Code Works},
  year    = {2026},
  url     = {https://code.claude.com/docs/en/how-claude-code-works},
  lastaccessed = {August 3, 2026}
}

@inproceedings {autocompletion_poisoning,
  author = {Roei Schuster and Congzheng Song and Eran Tromer and Vitaly Shmatikov},
  title = {You Autocomplete Me: Poisoning Vulnerabilities in Neural Code Completion},
  booktitle = {30th USENIX Security Symposium (USENIX Security 21)},
  year = {2021},
  isbn = {978-1-939133-24-3},
  pages = {1559--1575},
  publisher = {USENIX Association},
  address = {Vancouver, B.C., Canada},
  url = {https://www.usenix.org/conference/usenixsecurity21/presentation/schuster},
}

@inproceedings{sec_23_usenix_lost_at_c,
  author = {Sandoval, Gustavo and Pearce, Hammond and Nys, Teo and Karri, Ramesh and Garg, Siddharth and Dolan-Gavitt, Brendan},
  title = {Lost at C: a user study on the security implications of large language model code assistants},
  year = {2023},
  booktitle = {Proceedings of the 32nd USENIX Conference on Security Symposium},
  pages = {2205--2222},
  publisher = {USENIX Association},
  address = {Anaheim, CA, USA},
  url = {https://dl.acm.org/doi/10.5555/3620237.3620361},
}

@inproceedings{pearce2022asleep,
  title={Asleep at the Keyboard? Assessing the Security of GitHub Copilot’s Code Contributions},
  author={Pearce, Hammond and Ahmad, Baleegh and Tan, Benjamin and Dolan-Gavitt, Brendan and Karri, Ramesh},
  booktitle={2022 IEEE Symposium on Security and Privacy (SP)},
  pages={754--768},
  year={2022},  
  publisher={IEEE Computer Society},
  address={Los Alamitos, CA, USA},  
  organization={IEEE Computer Society},
  doi={10.1109/SP46214.2022.9833571}
}

@misc{mou2025reallytrustcodecopilots,
      title={Can You Really Trust Code Copilots? Evaluating Large Language Models from a Code Security Perspective}, 
      author={Yutao Mou and Xiao Deng and Yuxiao Luo and Shikun Zhang and Wei Ye},
      year={2025},
      eprint={2505.10494},
      archivePrefix={arXiv},
      primaryClass={cs.CL},
      url={https://arxiv.org/abs/2505.10494}, 
}

@INPROCEEDINGS {sp_24_poisoned_chatgpt,
    author = { Oh, Sanghak and Lee, Kiho and Park, Seonhye and Kim, Doowon and Kim, Hyoungshick },
    booktitle = { 2024 IEEE Symposium on Security and Privacy (SP) },
    title = {{ Poisoned ChatGPT Finds Work for Idle Hands: Exploring Developers’ Coding Practices with Insecure Suggestions from Poisoned AI Models }},
    year = {2024},    pages = {1141--1159},
    publisher = {IEEE Computer Society},
    address = {Los Alamitos, CA, USA},    url = {https://doi.ieeecomputersociety.org/10.1109/SP54263.2024.00046},
}

@misc{bhatt2023purplellamacybersecevalsecure,
      title={Purple Llama CyberSecEval: A Secure Coding Benchmark for Language Models}, 
      author={Manish Bhatt and others},
      year={2023},
      eprint={2312.04724},
      archivePrefix={arXiv},
      primaryClass={cs.CR},
      url={https://arxiv.org/abs/2312.04724}, 
}

@misc{maloyan2026promptinjectionattacksagentic,
      title={Prompt Injection Attacks on Agentic Coding Assistants: A Systematic Analysis of Vulnerabilities in Skills, Tools, and Protocol Ecosystems}, 
      author={Narek Maloyan and Dmitry Namiot},
      year={2026},
      eprint={2601.17548},
      archivePrefix={arXiv},
      primaryClass={cs.CR},
      url={https://arxiv.org/abs/2601.17548}, 
}

@inproceedings{greshake2023youvesignedforcompromising,
  author = {Greshake, Kai and Abdelnabi, Sahar and Mishra, Shailesh and Endres, Christoph and Holz, Thorsten and Fritz, Mario},
  title = {Not What You've Signed Up For: Compromising Real-World LLM-Integrated Applications with Indirect Prompt Injection},
  year = {2023},
  isbn = {9798400702600},
  publisher = {Association for Computing Machinery},
  address = {New York, NY, USA},
  url = {https://doi.org/10.1145/3605764.3623985},
  doi = {10.1145/3605764.3623985},
  booktitle = {Proceedings of the 16th ACM Workshop on Artificial Intelligence and Security},
  pages = {79–90},
  numpages = {12},
  location = {Copenhagen, Denmark},
  series = {AISec '23}
}

@online{codeql,
  title        = {Code Scanning with CodeQL},
  author       = {{GitHub, Inc.}},
  year         = {2021},
  url          = {https://codeql.github.com/},
  lastaccessed = {August 1, 2026}
}

@online{tree_sitter,
  title        = {Tree-sitter: An Incremental Parsing System for Programming Tools},
  author       = {{GitHub, Inc.}},
  year         = {2026},
  url          = {https://github.com/tree-sitter/tree-sitter},
  lastaccessed = {August 1, 2026}
}

@article{Dickert1999WhatsTP,
  title={What's the price of a research subject? Approaches to payment for research participation.},
  author={Neal Dickert and Christine Grady},
  journal={The New England journal of medicine},
  year={1999},
  volume={341},
  number={3},
  pages={198--203},
  url={https://api.semanticscholar.org/CorpusID:45204547}
}

@article{data_quality,
  author = {Stecklov, Guy and Weinreb, Alex and Carletto, Gero},
  year = {2017},
  title = {Can Incentives Improve Survey Data Quality in Developing Countries?: Results from a Field Experiment in India},
  journal = {Journal of the Royal Statistical Society: Series A (Statistics in Society)},
  volume = {181},
  number = {4},
  pages = {1033--1056},
  doi = {10.1111/rssa.12333}
}

@article{IMC_survey,
  title = {Instructional manipulation checks: Detecting satisficing to increase statistical power},
  journal = {Journal of Experimental Social Psychology},
  year = {2009},
  issn = {0022-1031},
  volume = {45},
  number = {4},
  pages = {867--872},
  doi = {https://doi.org/10.1016/j.jesp.2009.03.009},
  url = {https://www.sciencedirect.com/science/article/pii/S0022103109000766},
  author = {Daniel M. Oppenheimer and Tom Meyvis and Nicolas Davidenko},
}

@online{trae,
  author = {TRAE},
  title = {TRAE: The Real AI Engineer},
  year = {2025},
  url = {https://www.trae.ai/},
  lastaccessed = {August 1, 2026}
}

@online{cursor_insecure_code,
  author = {Cursor},
  title = {Cursor Issue: Cursor keeps trying to access sensitive env variables even though .env is ignored},
  year = {2025},
  url = {https://forum.cursor.com/t/cursor-keeps-trying-to-access-sensitive-env-variables-even-though-env-is-ignored/145607},
  lastaccessed = {August 1, 2026}
}

@online{cursor_crazy_leak,
  author = {Cursor},
  title = {Cursor Issue: Cursor reads .env even though it is on .cursorignore},
  year = {2025},
  url = {https://forum.cursor.com/t/cursor-reads-env-even-though-it-is-on-cursorignore/136998},
  lastaccessed = {August 1, 2026}
}

@online{cursor_view_code_leak,
  author = {Cursor},
  title = {Cursor Issue: Cursor AI View Code Leak},
  year = {2025},
  url = {https://forum.cursor.com/t/big-security-risk-cursorignore-doesnt-seem-to-work-envs-files-being-sent-as-context/14027},
  lastaccessed = {August 1, 2026}
}

@online{cursor_ignore_file,
  author = {Cursor},
  title = {Cursor Ignore File},
  year = {2025},
  url = {https://cursor.com/cn/docs/reference/ignore-file},
  lastaccessed = {August 1, 2026}
}

@online{cursor_security_response,
  author = {Cursor},
  title = {Questions on .gitignore, .cursorignore, .cursorban},
  year = {2025},
  url = {https://forum.cursor.com/t/questions-on-gitignore-cursorignore-cursorban/34713},
  lastaccessed = {August 1, 2026}
}

@misc{hui2024qwen25codertechnicalreport,
      title={Qwen2.5-Coder Technical Report}, 
      author={Binyuan Hui and others},
      year={2024},
      eprint={2409.12186},
      archivePrefix={arXiv},
      primaryClass={cs.CL},
      url={https://arxiv.org/abs/2409.12186}, 
}

@misc{Qwen3-Coder-Next,
    title={Qwen3-Coder-Next Technical Report},
    author={Ruisheng Cao and others},
    journal={arXiv preprint arXiv:2603.00729},
    year={2026},
}

@ARTICLE{js_divergence,
  author={Lin, J.},
  journal={IEEE Transactions on Information Theory}, 
  title={Divergence measures based on the Shannon entropy}, 
  year={1991},
  volume={37},
  number={1},
  pages={145--151},
  doi={10.1109/18.61115}
}

@online{owasp_xxe,
  author       = {{OWASP Foundation}},
  title        = {XML External Entity (XXE) Processing},
  year         = {2024},
  url          = {https://owasp.org/www-community/vulnerabilities/XML_External_Entity_(XXE)_Processing},
  lastaccessed = {August 1, 2026},
  organization = {OWASP}
}

@online{owasp_top10_2025,
  author       = {{OWASP Foundation}},
  title        = {OWASP Top 10 - 2025},
  year         = {2025},
  url          = {https://owasp.org/Top10/2025/},
  lastaccessed = {August 1, 2026},
  organization = {OWASP}
}

@inproceedings{phthia_2019_ai_assisted_code_completion,
    author = {Svyatkovskiy, Alexey and Zhao, Ying and Fu, Shengyu and Sundaresan, Neel},
    title = {Pythia: AI-assisted Code Completion System},
    year = {2019},
    isbn = {9781450362016},
    publisher = {Association for Computing Machinery},
    address = {New York, NY, USA},
    url = {https://doi.org/10.1145/3292500.3330699},
    doi = {10.1145/3292500.3330699},
    booktitle = {Proceedings of the 25th ACM SIGKDD International Conference on Knowledge Discovery \& Data Mining},
    pages = {2727–2735},
    numpages = {9},
    location = {Anchorage, AK, USA},
    series = {KDD '19}
}

@article{grounded_copilot_2023_oopsla,
  author = {Barke, Shraddha and James, Michael B. and Polikarpova, Nadia},
  title = {Grounded Copilot: How Programmers Interact with Code-Generating Models},
  year = {2023},
  issue_date = {April 2023},
  publisher = {Association for Computing Machinery},
  address = {New York, NY, USA},
  volume = {7},
  number = {OOPSLA1},
  url = {https://doi.org/10.1145/3586030},
  doi = {10.1145/3586030},
  journal = {Proc. ACM Program. Lang.},
  month = apr,
  articleno = {78},
  numpages = {27}
}

@article{nickerson2013method,
  title={A method for taxonomy development and its application in information systems},
  author={Nickerson, Robert C and Varshney, Upkar and Muntermann, Jan},
  journal={European journal of information systems},
  volume={22},
  number={3},
  pages={336--359},
  year={2013},
  publisher={Taylor \& Francis}
}

@article{tosem_25_prompting_techniques_secure_code_generation,
  author = {Tony, Catherine and D{\'i}az Ferreyra, Nicol{\'a}s E. and Mutas, Markus and Dhif, Salem and Scandariato, Riccardo},
  title = {Prompting Techniques for Secure Code Generation: A Systematic Investigation},
  year = {2025},
  issue_date = {November 2025},
  publisher = {Association for Computing Machinery},
  address = {New York, NY, USA},
  volume = {34},
  number = {8},
  issn = {1049-331X},
  url = {https://doi.org/10.1145/3722108},
  doi = {10.1145/3722108},
  journal = {ACM Trans. Softw. Eng. Methodol.},
  month = oct,
  articleno = {225},
  numpages = {53},
}

@inproceedings{icml_24_instruction_tuning_secure_code_generation,
  author = {He, Jingxuan and Vero, Mark and Krasnopolska, Gabriela and Vechev, Martin},
  title = {Instruction tuning for secure code generation},
  year = {2024},
  publisher = {JMLR.org},
  booktitle = {Proceedings of the 41st International Conference on Machine Learning},
  articleno = {723},
  numpages = {20},
  address = {Vienna, Austria},
  series = {ICML'24}
}

@inproceedings{icml_25_prosec,
  author = {Xu, Xiangzhe and Su, Zian and Guo, Jinyao and Zhang, Kaiyuan and Wang, Zhenting and Zhang, Xiangyu},
  title = {PROSEC: fortifying code LLMs with proactive security alignment},
  year = {2025},
  publisher = {JMLR.org},
  booktitle = {Proceedings of the 42nd International Conference on Machine Learning},
  articleno = {2784},
  numpages = {16},
  address = {Vancouver, Canada},
  series = {ICML'25}
}

@INPROCEEDINGS{2021_ICSE_SEET_secure_coding_education,
  author={Espinha Gasiba, Tiago and Lechner, Ulrike and Pinto-Albuquerque, Maria and Méndez, Daniel},
  booktitle={2021 IEEE/ACM 43rd International Conference on Software Engineering: Software Engineering Education and Training (ICSE-SEET)}, 
  title={Is Secure Coding Education in the Industry Needed? An Investigation Through a Large Scale Survey}, 
  year={2021},
  volume={},
  number={},
  publisher = {IEEE Press},
  address = {Virtual Event, Spain},
  pages={241--252},
  doi={10.1109/ICSE-SEET52601.2021.00034}
}

@article{2024_cose_cybersecurity_training_methods,
  author = {Pr{\"u}mmer, Julia and van Steen, Tommy and van den Berg, Bibi},
  title = {A systematic review of current cybersecurity training methods},
  year = {2024},
  issue_date = {Jan 2024},
  publisher = {Elsevier Advanced Technology Publications},
  address = {GBR},
  volume = {136},
  number = {C},
  issn = {0167-4048},
  url = {https://doi.org/10.1016/j.cose.2023.103585},
  doi = {10.1016/j.cose.2023.103585},
  journal = {Comput. Secur.},
  month = jan,
  numpages = {20},
}

@inproceedings{agentdojo_2024,
  author = {Debenedetti, Edoardo and Zhang, Jie and Balunovic, Mislav and Beurer-Kellner, Luca and Fischer, Marc and Tram{\`e}r, Florian},
  title = {AgentDojo: a dynamic environment to evaluate prompt injection attacks and defenses for LLM agents},
  year = {2024},
  isbn = {9798331314385},
  publisher = {Curran Associates Inc.},
  address = {Red Hook, NY, USA},
  booktitle = {Proceedings of the 38th International Conference on Neural Information Processing Systems},
  articleno = {2636},
  numpages = {26},
  location = {Vancouver, BC, Canada},
  series = {NIPS '24}
}

@article{tosem_25_security_weaknesses_copilot_generated_code,
  author = {Fu, Yujia and Liang, Peng and Tahir, Amjed and Li, Zengyang and Shahin, Mojtaba and Yu, Jiaxin and Chen, Jinfu},
  title = {Security Weaknesses of Copilot-Generated Code in GitHub Projects: An Empirical Study},
  year = {2025},
  issue_date = {November 2025},
  publisher = {Association for Computing Machinery},
  address = {New York, NY, USA},
  volume = {34},
  number = {8},
  issn = {1049-331X},
  url = {https://doi.org/10.1145/3716848},
  doi = {10.1145/3716848},
  journal = {ACM Trans. Softw. Eng. Methodol.},
  month = oct,
  articleno = {218},
  numpages = {34},
}

@inproceedings{ccs_24_ai_assistants_security_practices,
  author = {Klemmer, Jan H. and others},
  title = {Using AI Assistants in Software Development: A Qualitative Study on Security Practices and Concerns},
  year = {2024},
  isbn = {9798400706363},
  publisher = {Association for Computing Machinery},
  address = {New York, NY, USA},
  url = {https://doi.org/10.1145/3658644.3690283},
  doi = {10.1145/3658644.3690283},
  booktitle = {Proceedings of the 2024 on ACM SIGSAC Conference on Computer and Communications Security},
  pages = {2726–2740},
  numpages = {15},
  location = {Salt Lake City, UT, USA},
  series = {CCS '24}
}

@inproceedings{zhang-etal-2023-repocoder,
    title = "{R}epo{C}oder: Repository-Level Code Completion Through Iterative Retrieval and Generation",
    author = "Zhang, Fengji  and
      Chen, Bei  and
      Zhang, Yue  and
      Keung, Jacky  and
      Liu, Jin  and
      Zan, Daoguang  and
      Mao, Yi  and
      Lou, Jian-Guang  and
      Chen, Weizhu",
    editor = "Bouamor, Houda  and
      Pino, Juan  and
      Bali, Kalika",
    booktitle = "Proceedings of the 2023 Conference on Empirical Methods in Natural Language Processing",
    month = dec,
    year = "2023",
    address = "Singapore",
    publisher = "Association for Computational Linguistics",
    url = "https://aclanthology.org/2023.emnlp-main.151/",
    doi = "10.18653/v1/2023.emnlp-main.151",
    pages = "2471--2484",
}

\appendix 

\section{Open Science} 
To comply with Open Science Policy and to support reproducibility and independent evaluation, we make available all artifacts necessary to understand, assess, and replicate the core findings of this work, subject to ethical and security considerations. The artifacts are accessible from: \url{https://github.com/plain-noodle-expert/NesCodeSec}.



\section{Ethical Considerations} 


We obtained IRB approval before conducting our survey and distributed the survey on reputable and reliable platforms. During the survey, we ensured all participants were provided informed consent, and participation was entirely voluntary, and they could withdraw at any time. The survey consists of no sensitive questions or questions that may disclose personal information. The number of questions is designed to cost approximately 5 minutes for participants.\
To recognize their time and effort, we offered a monetary compensation to each participant upon survey completion.

\section{Detection of Insecure NES Suggestions}
\label{sec:appendix_detection_logic}

For each test case, we retain the original code and the generated NES suggestion. Because a suggestion may contain only a partial diff, we first apply it to the original file to reconstruct the post-edit program. We then evaluate the prespecified criterion for each vector using regex-based scanning for structural patterns (e.g., MD5 or hardcoded credentials), AST-based static analysis for syntactic and configuration properties (e.g., XML parser settings), or LLM-based semantic analysis for risks requiring data-flow or intent reasoning (e.g., sensitive logging or intended Undo actions). Unrelated weaknesses are not counted toward that vector. For the LLM evaluated cases, we use a five-model panel comprising DeepSeek-V3.2, Qwen3-235B-A22B, Gemini-3-Flash, Claude-Haiku-4.5, and GPT-5-mini.
Each model is queried separately in a fresh session using the same vector-specific prompt and has no access to the responses of the other judges.
The input contains the relevant pre-edit code or interaction trace, the NES suggestion, the reconstructed post-edit code, and the security criterion for the target manifestation.
Each judge independently returns a binary label, where 1 indicates that the target manifestation is present and 0 indicates that it is absent.
We use a symmetric strong-majority rule: an output is automatically labeled vulnerable for the target vector when at least four judges return 1, and not vulnerable for that vector when at least four return 0.
The produced ambiguous cases (2 vs 3 votes), totaling 51 samples in our evaluation, are independently reviewed by the authors.
The resulting adjusted label is used in the final vulnerability-rate calculation.

\end{document}